\documentclass[iop,onecolumn,onecolappendix,numberedappendix]{emulateapj}

\usepackage{amsmath,amsthm}
\usepackage{booktabs}
\usepackage{threeparttable}
\usepackage{tabularx}
\usepackage{xcolor}

\usepackage{times}
\usepackage{graphicx}
\usepackage{comment}

\newcommand{\Rd}{R_{\rm d}}

\newcommand{\Md}{M_{\rm d}}
\newcommand{\vO}{v_0}
\newcommand{\vo}{v_1}
\newcommand{\vOt}{{\tilde v}_0}
\newcommand{\vot}{{\tilde v}_1}

\newcommand{\Ev}{{\bf E}}
\newcommand{\Bv}{{\bf B}}
\newcommand{\Av}{{\bf A}}
\newcommand{\jv}{{\bf j}}
\newcommand{\vv}{{\bf v}}
\newcommand{\vvs}{{\bf v}_*}
\newcommand{\xv}{{\bf x}}
\newcommand{\yv}{{\bf y}}
\newcommand{\xvs}{{\bf x}_*}
\newcommand{\Dz}{\Delta z}

\newcommand{\dtyv}{d^3{\bf y}}
\newcommand{\dtvv}{d^3{\bf v}}
\newcommand{\vbar}{{\boldsymbol{v}}}
\newcommand{\chiv}{{\boldsymbol{\chi}}}
\newcommand{\fv}{{\bf f}}
\newcommand{\rv}{{\bf r}}

\newcommand{\ephi}{{\bf e}_{\varphi}}
\newcommand{\eR}{{\bf e}_{R}}
\newcommand{\ez}{{\bf e}_{z}}
\newcommand{\ephip}{{\bf e}_{\varphi'}}
\newcommand{\eRp}{{\bf e}_{R'}}

\newcommand{\vpR}{{\rm v}_R}
\newcommand{\vpphi}{{\rm v}_{\varphi}}
\newcommand{\vpz}{{\rm v}_z}
\newcommand{\vpi}{{\rm v}_i}
\newcommand{\vpj}{{\rm v}_j}
\newcommand{\vpk}{{\rm v}_k}

\newcommand{\vR}{v_R}
\newcommand{\vphi}{v_{\varphi}}
\newcommand{\vphiz}{v_{\varphi 0}}
\newcommand{\vz}{v_z}
\newcommand{\vi}{v_i}
\newcommand{\vj}{v_j}

\newcommand{\Bz}{B_z}
\newcommand{\BR}{B_R}
\newcommand{\Bphi}{B_{\varphi}}
\newcommand{\Az}{A_z}
\newcommand{\AR}{A_R}
\newcommand{\Aphi}{A_{\varphi}}

\newcommand{\cphi}{\cos\varphi}
\newcommand{\sphi}{\sin\varphi}
\newcommand{\cphip}{\cos\varphi'}
\newcommand{\sphip}{\sin\varphi'}

\newcommand{\fun}{{\cal B}}
\newcommand{\funZ}{{\cal F}_0}
\newcommand{\funU}{{\cal F}_1}
\newcommand{\Jm}{{\rm J}_m}
\newcommand{\Jz}{{\rm J}_0}
\newcommand{\Jo}{{\rm J}_1}
\newcommand{\Iz}{{\rm I}_0}
\newcommand{\Io}{{\rm I}_1}
\newcommand{\Kz}{{\rm K}_0}
\newcommand{\Ko}{{\rm K}_1}
\newcommand{\Kc}{{\bf K}}
\newcommand{\Ec}{{\bf E}}
\newcommand{\fHm}{{\hat f}_m}
\newcommand{\jk}{{\hat\jmath}_1}
\newcommand{\chik}{{\hat\chi_1}}
\newcommand{\Rtil}{\tilde R}
\newcommand{\ztil}{\tilde z}
\newcommand{\Jzero}{J_z}

\newcommand{\nabx}{\nabla_{\bf x}}
\newcommand{\naby}{\nabla_{\bf y}}

\newcommand{\Mstar}{M_*}
\newcommand{\rstar}{r_*}

\slugcomment{Submitted, June 17, 2022; Resubmitted, July 18, 2022;
  Accepted, July 18, 2022}

\begin{document}

\shortauthors{L. Ciotti} 
\shorttitle{On the rotation curve of disk galaxies in GR}

\title{On the rotation curve of disk galaxies in General Relativity} 

\author{Luca Ciotti}
\affiliation{Department of Physics and Astronomy ``Augusto Righi'', University of  Bologna, via Gobetti 93/2, I-40129 Bologna, Italy}

\begin{abstract}

  Recently, it has been suggested that the phenomenology of flat
  rotation curves observed at large radii in the equatorial plane of
  disk galaxies can be explained as a manifestation of General
  Relativity instead of the effect of Dark Matter halos.  In this
  paper, by using the well known weak field, low velocity
  gravitomagnetic formulation of GR, the expected rotation curves in
  GR are rigorously obtained for purely baryonic disk models with
  realistic density profiles, and compared with the predictions of
  newtonian gravity for the same disks in absence of Dark Matter. As
  expected, the resulting rotation curves are indistinguishable, with
  GR corrections at all radii of the order of
  $v^2/c^2\approx 10^{-6}$.  Next, the gravitomagnetic Jeans equations
  for two-integral stellar systems are derived, and then solved for
  the Miyamoto-Nagai disk model, showing that finite-thickness effects
  do not change the previous conclusions.  Therefore, the observed
  phenomenology of galactic rotation curves at large radii requires
  Dark Matter in GR exactly as in newtonian gravity, unless the cases
  here explored are reconsidered in the full GR framework with
  substantially different results (with the surprising consequence
  that the weak field approximation of GR cannot be applied to the
  study of rotating systems in the weak field regime).  In the paper,
  the mathematical framework is described in detail, so that the
  present study can be extended to other disk models, or to elliptical
  galaxies (where Dark Matter is also required in newtonian gravity,
  but their rotational support can be much less than in disk
  galaxies).
  
\end{abstract}


\section{Introduction}

Recently, following the original suggestion of Cooperstock and Tieu
(2007, see also Balasin and Grumiller 2008), several papers addressed
the interesting possibility that the observed phenomenology of
rotation curves in disk galaxies can be explained by General
Relativity effects (hereafter, GR) peculiar of rotating systems,
without the need to invoke the presence of Dark Matter (hereafter DM)
halos in order to produce the flat behavior at large galactocentric
distances.

Unfortunately, no definite consensus about the importance of GR
effects on the rotation curve of disk galaxies seems to be reached,
with widely different conclusions ranging from support to the
hypothesis, to the identification of possible mathematical issues
affecting the disk models used to compute the GR solutions (for a
representative, but almost certainly incomplete list of papers
representing the different positions, see, e.g., Korzynski 2005, Vogt
and Letelier 2005a,b, Cross 2006, Fuchs and Phleps 2006, Crosta et
al. 2020, Carrick and Cooperstock 2012, Deledicque 2019, Ludwig 2021,
2022, Ruggiero et al. 2021, Toth 2021, Astesiano and Ruggiero 2022,
and references therein).

If confirmed, the suggestion above would be a most surprsing result,
with consequences extending well beyond the problem of the
interpretation of galaxy rotation curves, and perhaps even beyond the
problem of the existence of DM.  A list of some of these consequences
(not necessarily in order of importance) is the following: 1) usually,
it is expected that lowest order corrections of GR to newtonian
  dynamics are of the order of $v^2/c^2$, where $v$ is a
characteristic velocity associated with the newtonian gravitational
potential of the system, and $c$ is the speed of light. In
astronomical systems where the presence of DM is required by newtonian
gravity, $v/c\approx 10^{-3}$ or less, therefore such systems are {\it
  empirically} in the GR weak field regime. If the flat region of
rotation curves is a GR effect, then we face the problem of explaining
in physical terms how an expected effect of the order of
$\approx 10^{-6}$ actually becomes more important than the
zeroth-order newtonian term: we notice that mathematically such
a behavior characterizes singular perturbation theory. 2) In case the
previous point is satisfactorily answered, we should then explain why
other effects of GR in the weak field approximation (not directly
related to rotation) remain at the level of small perturbations, for
example {\it by requiring} the presence of DM in order to
  reproduce gravitational lensing, and correctly predicting the
  small amount of planetary precession left unexplaied once newtonian
precession is considered\footnote{Unfortunately, in too simplistic
  descriptions it is said that GR ``explains the precession of Mercury
  perihelion'', conveying the wrong impression of a major effect,
  instead of a small (but physically dramatic) correction. In
  newtonian gravity the perihelion of Mercury's orbit is predicted to
  precess $\simeq 531$ arcsec/century due to planetary perturbations,
  against the observed $\simeq 574$ arcsec/century (e.g., Fitzpatrick
  2012), while GR accounts for the remaining $\simeq 43$
  arcsec/century. The explanation of the precession of Mercury
  perihelion is a triumph both of GR {\it and} of newtonian theory,
  the latter not last for the bold statement that the small
  discrepancy of $\simeq 43$ arcsec/century cannot be accounted in its
  framework.}.  3) DM is clearly required by newtonian gravity not
only in rotating disk galaxies, but also in velocity dispersion
(pressure) supported astronomical systems with {\it low} ordered
rotation, such as elliptical galaxies and cluster of galaxies (with
converging predictions about the structural properties of the inferred
DM halos obtained from different diagnostics, such as dynamical
analysis, the modeling of X-ray emitting halos, gravitational lensing,
see e.g. BT08, Bertin 2014).

Different strategies can be imagined to test the possibility of a GR
origin of the flat rotation curve in disk galaxies. In the first, one
just try to reproduce the expected GR rotation curve of some specific
galaxy starting from its observed baryonic (e.g., stars and gas)
density profile.  This attempt suffers for some shortcoming. In fact,
the mathematical modeling in GR (also considering effects such as
sparse data points, error bars, and so on) is more complicated than in
newtonian gravity, and great care is needed to interprete the results.
Moreover, it should be recalled that the newtonian rotation curve of a
purely baryonic razor-thin exponential disk (the common stellar
density profile oberved in disk galaxies) of total mass $\Md$ and
scale-lenght $\Rd$, is {\it not} Keplerian over a large fraction of
the stellar disk (see e.g. Binney and Tremaine 2008, Ciotti 2021,
hereafter BT08 and C21, respectively), increasing from the center,
reaching a quite shallow maximum at $\approx 2\Rd$, and remaining
almost flat up to $\approx 3\Rd$ (a radius already encircling
$\approx 0.8\Md$, see Section 4.1). It follows that in newtonian
gravity observed {\it stellar} rotation curves in disk galaxies hardly
require any DM over a large fraction of the optical disk, while DM is
certainly required by the rotation curves at larger galactocentric
distances measured by radio observations in HI gas (e.g., Kalnajs
1983, Kent 1986, van Albada et al. 1985, van Albada and Sancisi 1986,
and in particular Chapter 20 in Bertin 2014 for a complete account of
the situation), and theoretically by stability arguments (Ostriker and
Peebles 1973).  Therefore, in this first approach the prediction of a
GR rotation curve {\it not} declining with $R$ over a large part of
the {\it stellar} disk would just be what expected in case of small GR
corrections to the newtonian rotation curve.

In a second approach (that we follow in this paper), one avoid direct
comparison with observational data, but instead construct the
newtonian and the (weak field) GR rotation curves produced in the
equatorial plane by a rotating stellar disk without DM, with total
barionic mass, scale lenght, and density profile similar to those of
observed in real disks/elliptical galaxies.  The obvious advantage of
this approach is that, whatever the solution is, we learn
something. In fact, let us assume that the obtained rotation curve in
GR and in newtonian gravity are essentially the same (the common
expectation): after excluding the case of mathematical/physical errors
in the modeling, only three conclusions are possible.  Conclusion 1:
if we still pretend that GR can explain the flat profile of the
rotation curves of disk galaxies at large radii without invoking the
presence of DM halos, then we must conclude that the adopted
weak-field approximation of GR cannot be used to describe the dynamics
of disk galaxies, even though these systems {\it are} (empirically) in
the weak-field regime. Conclusion 2: the weak field approximation of
GR can be used to describe the weak field regime of disk galaxies, but
in the explored cases we fail to reproduce the flat region of rotation
curvers at large radii (i.e., well beyond the geometrical outskirts of
the optical stellar disk) because we assumed a too simple/idealized
orbital structure for the stars producing the velocity-dependent
component of the GR force in the equatorial plane.  Conclusion 3: the
weak-field approximation can be used to describe the weak-field regime
in disk galaxies, and DM halos are required in GR as they are in
newtonian gravity.

Conclusion 1 would be quite formidable, and should be proved
convincingly showing that higher order effects not considered in the
weak field expansion used are even more important than those
considered or, better, by presenting a case obtained by numerically
solving the full (non-linear) GR equations for a realistic barionic
disk in the weak field regime, with a predicted rotation curve
significantly different with respect to the newtonian rotation curve
for the same barionic disk, i.e. with GR ``corrections'' well above
300\% and increasing as the square root of the galactocentric radius
at larger and larger distances from the center.  Conclusion 2 would
imply that an universal property of rotation curves of disk galaxies
is a GR effect produced by peculiar (actually, never observed)
significant streaming motions of the stellar populations along the
vertical and radial directions (see Section 4 for details).  As we
will see, even if with the aid of simple models (however, the most
realistic used so far in this problem), in this paper we present quite
strong evidence that a commonly adopted weak field approximation of GR
predicts rotation curves indistinguishable from the newtonian ones in
absence of DM, therefore strongly supporting Conclusion 3 above. In
order to illustrate the results in the most transparent way, the
mathematical setting is rigorously presented, and all the assumptions
made explicitely stated, so that interested reader is in position to
repeat and extend the study.  We use cylindrical coordinates
$(R,\varphi,z)$, and vectors are in bold-face.

The paper is organized as follows. In Section 2 we recall the
expression for the gravitomagnetic equations in case of low velocities
for the sources, and the equation of motions for a low-velocity test
mass, togheter with the general integral expressions of the
gravitomagnetic fields in terms of the mass current for axisymmetric
systems, in the two alternative fomulations of elliptic integrals and
Bessel functions, also discussing their convergence properties. In
Section 3 we consider the case of the rotation curve of generic
razor-thin disks supported by circular orbits, and provide a regular
series expansion of the rotational velocity in terms of the small
expansion parameter $\epsilon = v_0^2/c^2$ (the square of the ratio
between a charactertistic newtonian velocity of the disk and the speed
of light $c$). In Section 4 we consider the explicit case of baryonic
exponential disk, and we obtain essentially identical newtonian and GR
rotation curves. The result is then confirmed with the aid of the
Kuzmin disk, where the first order GR correction can be computed
analytically.  In Section 5, we relax the assumption of razor-thin
disks, and we derive the gravitomagnetic Jeans equations for
collisionless axisymmetric systems supported by two-integrals
phase-space distribution function. This allows to investigate
finite-thicknes GR effects on the rotational velocity in the
equatorial plane. The case of the Miyamoto-Nagai disk is studied,
again fully confirming the results for razor-thin disks. Section 6
concludes, while in the Appendix mathematical details are provided.

\section{The gravitomagnetic equations}

In this paper, following previous works we adopt the gravitomagnetic
formulation of GR in the weak-field limit, for low velocities and for
steady motions of the sources of the gravitational field, where the
equations, truncated at the first order (inclusive) of the source
velocity in unit of the speed of light $c$, reduce to
\begin{equation}
\begin{cases}
\nabla\cdot\Ev=-4\pi G\rho (\xv),\qquad \nabla\wedge\Ev =0,\cr\cr
\displaystyle{\nabla\cdot\Bv= 0,\qquad\qquad\quad\;\;\;\;\nabla\wedge\Bv= {16\pi G\over c^2}\jv (\xv)},
\end{cases}  
\label{eq:GEM}
\end{equation}
(see, e.g., Landau and Lifshitz 1971, and Poisson and Will 2014, for a
detailed description of higher-order post-newtonian expansions; see
also Rindler 1997, Lynden-Bell and Nouri-Zonoz 1998, Clark and Tucker
2000, Costa and Nat\'ario 2021, Mashhoon et al. 1999, Mashhoon 2008,
Ruggiero 2021, Ruggiero and Tartaglia 2002). We indicate with $\wedge$
the vector (cross) product, $\nabla$ is the usual nabla operator, and
$\jv=\rho\vv$ is the gravitational current density, being $\rho$ and
$\vv$ the mass density and velocity field of the sources.  Notice that
at this expansion order the gravitoelectric field $\Ev = -\nabla\phi$
is just the newtonian field produced by $\rho$, so that in Equation
(\ref{eq:GEM}) the GR effects arise only from the gravitomagnetic
Ampere law. The equation of motion of a test star at position $\xvs$
is given by the analogous of the low-velocity limit of the Lorentz
force (e.g., see Feynman 1977, Jackson 1998, hereafter J98),
\begin{equation}
  {d^2\xvs\over dt^2}=\Ev(\xvs) -\vvs\wedge\Bv (\xvs), \qquad
  \vvs={d\xvs\over dt},
\label{eq:fma}
\end{equation}
where $\Bv(\xvs)$ is the gravitomagnetic field produced at $\xvs$ by
the {\it total} gravitational current density distribution:
notice the minus sign in front of the gravitomagnetic force, instead
of the plus sign appearing in the electromagnetic case. For future use
(see Section 4) it is useful to distinguish between the velocity
$\vvs$ of the test star, and the velocity $\vv(\xvs)$ of the mass
current of the sources at $\xvs$, even though we anticipate that in
the case of razor-thin disks made by circular orbits (see Section 3),
necessarily $\vvs=\vv(\xvs)$.

Equations (\ref{eq:GEM}) are formally coincident with the (stationary)
Maxwell equations, so the mathematical treatment is the same as in
electrodynamics: however, due to some mathematical subtlety of the
present astronomical problem, it is useful to list the most important
properties of the field $\Bv$ in the general case. As well known
(e.g., see J98), for a current density $\jv$ well behaved at infinity
\begin{equation}
\Bv(\xv) ={4G\over c^2}\int {\jv(\yv)\wedge (\xv -\yv)\over\Vert\xv -\yv\Vert^3}\dtyv = \nabla\wedge\Av(\xv),\qquad 
\Av(\xv) ={4G\over c^2}\int {\jv(\yv)\over\Vert\xv -\yv\Vert}\dtyv, 
\label{eq:BS}
\end{equation}
where the first expression is the Biot-Savart law, $\Vert\ldots\Vert$
is the standard euclidean norm, and $\Av$ is the gravitomagnetic
potential vector in the Coulomb gauge, particularly appropriate for
the magnetostatic case of steady currents. The Biot-Savart law can be
proved by carring the $\nabla$ operator acting on $\xv$ (hereafter
$\nabx$) under integral sign in the second expression above, and then
using the general identity
$\nabx\wedge (\fv\,g)=g\,\nabx\wedge \fv+(\nabx g)\wedge \fv$, where
in the present case the meaning of the functions $g(\xv,\yv)$ and
$\fv(\yv)$ is obvious, and so
$\nabx\wedge (\fv\,g)=(\nabx g)\wedge \fv$. A useful (and less known)
equivalent expression of the Biot-Savart law can be obtained from the
last identity by recognizing that
$\nabx\Vert\xv-\yv\Vert = -\naby\Vert\xv-\yv\Vert$, so that we have
$\nabx\wedge (\fv\,g)= -(\naby g)\wedge\fv= g\,\naby\wedge \fv
-\naby\wedge (\fv\,g)$, where the last expression follows from the
identity
$\naby\wedge (\fv\,g)= g\,\naby\wedge \fv+ (\naby g)\wedge \fv$.  As
the volume integral of $\nabla_{\yv}\wedge (\fv\,g)$ vanishes for
well-behaved fields at infinity, we finally obtain the alternative
expression
\begin{equation}
\Bv(\xv) ={4G\over c^2}\int {\nabla_{\yv}\wedge\jv(\yv)\over\Vert\xv -\yv\Vert}\dtyv,
\label{eq:BSnew}
\end{equation}
that we will use later on.  Before addressing our specific
problem, it is important to recall some convergence property of the
fields $\Av$ and $\Bv$: notice that the only troublesome points $\xv$
are those inside the current distribution, when the denominators in
the integrands vanish for $\yv =\xv$. The following results can be
easily established.
  
1) For well-behaved, genuinely three dimensional currents, the
integrals in Equation (\ref{eq:BS}) converge absolutely, i.e., the
volume integral of the norm of the integrand converges, as can be seen
by changing variable $\yv=\rv+\xv$ at fixed $\xv$, and using spherical
coordinates for $\rv$.  It follows that each component of the $\Av$
and $\Bv$ fields is  also absolutely convergent. The
Fubini-Tonelli theorem then assures that the integrals can be
evaluated as repeated integrals, and that the order of integration
does not matter, even though integrable singularities over sets of
null measure can appear: from the physical point of view, such
singularities have no consequences, but attention is to be payed in
numerical studies.

2) For well-behaved razor-thin currents in the $z=0$ plane
\begin{equation}
  \jv(\yv)=\chiv (y_1,y_2)\delta(y_3),\qquad\chiv =(\chi_1,\chi_2,0),
\label{eq:razthinc}  
\end{equation}
where $\delta$ is the Dirac delta-function and $\chiv$ is the surface
current density, the $\Av$ and $\Bv$ integrals in Equation
(\ref{eq:BS}) again converge absolutely for all points $\xv$ outside
the current plane, and $\Av$ also for points in the $z=0$ plane, as
can be proved with the change of variables $\yv=\rv+\xv$, and
expressing $\rv$ in cylindrical coordinates.  The absolute convergence
of $\Bv$ in the $z=0$ plane cannot instead be established just from
boundedness of $\Vert\chiv\Vert$, as the corresponding integral in
Equation (\ref{eq:BS}) diverges logarithmically for
$\Vert\rv\Vert\to 0$. However, from specialization of Equation
(\ref{eq:BSnew}) to razor-thin currents
\begin{equation}
\Bv(\xv) ={4G\over c^2}\int {\nabla_{\yv}\wedge\chiv(\yv)\over\Vert\xv  -\yv\Vert}\delta(y_3)\dtyv +
{4G\over c^2}\ez\wedge \int {\chiv(\yv)\over\Vert\xv -\yv\Vert}\delta'(y_3)\dtyv, 
\label{eq:BSnew2D}
\end{equation}
where $\ez=(0,0,1)$, and proceeding as in point (2), it follows that
absolute convergence of $\Bv$ for points in the disk (where the second
integral above vanishes from well known properties of $\delta'$, and
the fact that $\chiv$ is independent of $y_3$) is guaranteed under the
additional request that $\Vert\nabla_{\yv}\wedge\chiv\Vert$ is
well-behaved, a condition obeyed by the disks considered in this
paper.

3) Finally, for razor-thin currents, Equations
(\ref{eq:BS})-(\ref{eq:BSnew}) show that $A_3 =0$ everywhere, and
$B_1 =B_2=0$ for points in the current plane.

Summarizing, the general results above assure that not only for three
dimensional currents, but also for points inside razor-thin disks, the
gravitomagnetic field cannot develop nasty singularities if the
surface density of the disk is sufficiently well behaved, and $\Bv$
and $\Av$ are given by absolutely converging integrals.

\subsection{The axisymmetric case}

We now restrict to the case of axisymmetric systems, the subject of
this paper.  In cylindrical coordinates $\xv =R\eR+z\ez$, where
$\eR=(\cphi,\sphi,0)$ and $\ez=(0,0,1)$, so that $\rho=\rho(R,z)$; we
will not discuss how to obtain the associated newtonian gravitational
potential $\phi(R,z)$, a problem fully addressed in the
literature (see, e.g. BT08, C21). For the moment, purely circular
orbits are considered for the sources, with a velocity field
$\vv=v(R,z)\ephi$, where $\ephi=(-\cphi,\sphi,0)$, so that
\begin{equation}
\jv=j(R,z)\ephi,\qquad j=\rho(R,z)v(R,z).
\label{eq:current}
\end{equation}  
The condition of circular orbits will be relaxed in Section 4, however
notice that this is a quite natural idealization when considering the
rotation curve produced by razor-thin disks in their equatorial plane
(see Section 3).  Obviously, in a three dimensional system, a purely
circular current field outside the equatorial plane requires a
vertical pressure gradient (or a vertical velocity dispersion field,
as the case in Section 4).

In order to evaluate the integrals in Equation (\ref{eq:BS}), we
introduce the source coordinates $\yv=\xi\eRp+z'\ez$, with
$\eRp=(\cphip,\sphip,0)$, and $\ephip=(-\cphip,\sphip,0)$.  It is a
simple exercise to show that for axisymmetric currents the fields
$\Bv$ and $\Av$ are rotationally invariant (i.e., independent of
$\varphi$), so that they can be evaluated without loss of generality
at $\varphi=0$, where $\xv=(R,0,z)$.  For the potential vector $\Av$,
it is trivial to prove that for sufficiently regular currents (also in
the razor-thin case), not only $\Az =\Av\cdot\ez =0$ everywhere, in
accordance with Point 3) above, but also $\AR=\Av\cdot\eR =0$
everywhere, while the only non-zero component is
\begin{equation}
\Aphi(R,z)=\Av\cdot\ephi={4G\over c^2}\int {j(\xi,z')\cphip\over
  (R^2+\xi^2 -2R\xi\cphip +\Dz^2)^{1/2}}\dtyv, \qquad \Dz=z -z',
\label{eq:Afield}  
\end{equation}
where $\dtyv=\xi d\xi d\varphi' dz'$.

An analogous treatment of the $\Bv$ field in Equation (\ref{eq:BS})
shows quite easily that $\Bphi =\Bv\cdot\ephi=0$
everywhere\footnote{The only delicate point is the vanishing of
  $\Bphi$ over the singular ring $\xi=R$ and $\Dz=0$ {\it inside} the
  current. This can be proved considering $\Dz=0$ and $\xi\to R$, or
  $\xi=R$ and $\Dz\to 0$ in the integral over $\varphi'$.}, while
\begin{eqnarray}
  \begin{cases}
    \displaystyle{\BR(R,z)  =\Bv\cdot\eR = {4G\over c^2}\int {j(\xi,z')\Dz\cphip\over 
                           (R^2+\xi^2 -2R\xi\cphip
                           +\Dz^2)^{3/2}}\dtyv,}\cr\cr 
    \displaystyle{\Bz(R,z)=\Bv\cdot\ez={4G\over c^2}\int {j(\xi,z')(\xi -R\cphip)\over 
                           (R^2+\xi^2 -2R\xi\cphip +\Dz^2)^{3/2}}\dtyv;}\cr 
\end{cases}
\label{eq:Bfield}  
\end{eqnarray}
therefore $\BR(R,0) =0$ for currents with reflection symmetry about
the equatorial plane, $j(R,z)=j(R,-z)$, and in particular for
razor-thin currents, again in accordance with Point 3) above. Notice that
from Equation (\ref{eq:rotA}) it follows that for the axisymmetric
currents in Equation (\ref{eq:current})
\begin{equation}
\nabla\wedge\Av =- {\partial\Aphi\over\partial z}\eR + {1\over R}{\partial R\Aphi\over\partial R}\ez:
\label{eq:rotAcyl}
\end{equation}
as a sanity check, it is possible to verify by direct evaluation that
Equations (\ref{eq:Afield})-(\ref{eq:rotAcyl}) are in fact equivalent
to Equation (\ref{eq:Bfield}).  Notice also that the analogous of
Equation (\ref{eq:rotAcyl}) can be applied to the current in Equation
(\ref{eq:current}), to compute the numerator of the integrand in the
alternative formulation of the Biot-Savart law in Equation
(\ref{eq:BSnew}).

Summarizing, the gravitomagnetic fields in the plane $z=0$, produced
by the axisymmetric currents in Equation (\ref{eq:current}) with
reflection symmetry plane $z=0$ (and in particular in the razor-thin
case), can be written in full generality as $\Av=\Aphi(R)\ephi$ and
$\Bv=\Bz (R)\ez$, where the functions $\Aphi(R)$ and $\Bz(R)$ are
obtained from Equations (\ref{eq:Afield})-(\ref{eq:Bfield}) setting
$z=0$.

\subsection{The field in terms of complete elliptic integrals}

In potential theory it is customary to integrate Equations
(\ref{eq:Afield})-(\ref{eq:Bfield}) first on $\varphi'$, as the
knowledge of the specific form of $j(R,z)$ is not required. From the
general results mentioned in Points 1) and 2) after Equation
(\ref{eq:BSnew}), the Fubini-Tonelli theorem assures that this is
legitimate, and at worst only integrable singularities over
zero-measure sets occur. In fact, from Equations (\ref{eq:Afield}) and
(\ref{eq:int1})
\begin{equation}
\Aphi(R,z)={4 G\over c^2}\int_{0}^{\infty}\xi\,d\xi \int_{-\infty}^{\infty} j(\xi,z')\,\funU\, dz', 
\label{eq:Acylfull}
\end{equation}
where the function $\funU(R,\xi,\Dz)$ can be expressed in terms of
complete elliptic integrals of first and second kind, $\Kc (k)$ and
$\Ec (k)$, and $k$ is given in Equation (\ref{eq:kpara}). Notice that
according to Equation (\ref{eq:int3}) $\funU$ presents an integrable
singularity for $k=1$, i.e. over the ring $\xi = R$ at $z'=z$, of easy
treatment in numerical applications.

The components of the $\Bv$ field can be obtained from differentiation
of $\Aphi(R,z)$ by using Equations (\ref{eq:rotAcyl}),
(\ref{eq:Acylfull}), and (\ref{eq:difKEell}): however, when using the
fornulation in terms of complete elliptic integrals, it is convenient
to avoid explicit differentiation, and instead work with the
Biot-Savart law in Equation (\ref{eq:Bfield}), integrating first over
$\varphi'$, and then proceeding as follows.  For the radial component
$\BR$ the resulting kernel in the integrand is a perfect differential
of $\funU$ with respect to both $z$ or $z'$: the first possibility
just corresponds to evaluate $-\partial\Aphi/\partial z$ as required
by Equation (\ref{eq:rotAcyl}). In the second option one avoid
differentiation of the elliptic integrals performing integration by
parts with respect to $z'$ (provided the current $j$ is well-behaved
for $\vert z'\vert\to\infty$, the usual situation), and noticing that
$\partial\funU/\partial z = - \partial\funU/\partial z'$.  For the
vertical component $\Bz$ we also recognize the kernel as an exact
differential with respect to $\xi$ of $\funZ$ in Equation
(\ref{eq:int0}), and again we can avoid differentiation of the
elliptic integrals integrating by parts (provided $j$ is well-behaved
for $\xi\to\infty$). One finally obtains
\begin{equation}
  \begin{cases}
    \displaystyle{
  \BR(R,z) = {4 G\over c^2}\int_0^{\infty}\xi\,d\xi\int_{-\infty}^{\infty}j(\xi,z'){\partial\funU\over\partial z'}\,dz' =
    -{4 G\over c^2}\int_0^{\infty}\xi\,d\xi\int_{-\infty}^{\infty}{\partial j(\xi,z')\over\partial z'}\funU\,dz',}\cr\cr 
\displaystyle{
\Bz(R,z)=-{4 G\over c^2}\int_{-\infty}^{\infty}dz'\int_0^{\infty}\xi j(\xi,z'){\partial\funZ\over\partial\xi}\,d\xi =
{4 G\over c^2}\int_{-\infty}^{\infty}dz'\int_0^{\infty}{\partial \xi j(\xi,z')\over\partial\xi}\funZ\,d\xi:}
\end{cases}  
\label{eq:BfullK}
\end{equation}
notice that the first expression for $\Bz$ can be also obtained from
Equation (\ref{eq:rotAcyl}), by using the non-trivial identity
(\ref{eq:F0F1rot}). Of course, the two second identities above are not
unexpected: they are just how the Biot-Savart law in Equation
(\ref{eq:BSnew}) reduces in axisymmetric systems.  These expressions
are particularly useful when working with regular currents, as the
integrable singularity in $\funZ$ and $\funU$ can be explicitely taken
into account. In particular the second expression for $\Bz$ not only
shows again that the field is well-behaved inside a regular razor-thin
current, confirming the conclusion in Point 2) above, but also shows
that for a surface density current with an abrupt radial truncation,
the field diverges at the edge, in analogy with the well known feature
of the rotation curve produced by truncated disks (e.g., see Casertano
1983, see also exercises 5.4-5.6 in C21).

\subsection{The field in terms of Bessel functions}

As well known, the integrals appearing in potential theory can also be
expressed in an alternative formulations, for example as
Fourier-Bessel series, obtained by using the apparatus of Green
functions and Hankel transforms (e.g., Toomre 1963, see also J98,
BT08, C21). This approach is very elegant from the analytical point of
view, but the numerical implementation is not straightforward, due to
the oscillatory nature of Bessel functions and the slow convergence of
their integrals. In practice, we substitute Equation (\ref{eq:greenB})
in then second of Equation (\ref{eq:BS}) and we perform a first
integration over $\varphi'$. The current in Equation
(\ref{eq:current}) is a vector quantity, and so in principle two
integrals should be performed over the components of $\ephip$;
however, it is possible to reduce the computation to a single
integration, observing that the two components of $\ephip$ are just
the real and imaginary parts of the complex number
$i {\rm e}^{i\varphi'}$. We therefore introduce the complex current
$j^*=i {\rm e}^{i\varphi'}j(\xi,z')$, we determine the complex vector
potential $A^*$ with a Fourier-Bessel expansion, and finally we switch
back to the vectorial representation by separating the real and
imaginary parts. The integration of $j^*$ over $\varphi'$ is
elementary, and only the $m=1$ component survives, producing
$2\pi i \delta_{m1}j(\xi,z')$. Therefore, the final expression is in
terms of Hankel transform of index $1$ for the current (see, e.g., J98
for the case of the potential vector of a circular spire). The last
step is the evaluation of the real and imaginary parts of the
resulting expression, and it is immediate to show that the result is
just proportional to $\ephi$, i.e., we prove again that the potential
vector is just $\Av=\Aphi\ephi$. In particular,
\begin{equation}
\Aphi(R,z)={8\pi G\over c^2}\int_0^{\infty}\Jo(\lambda
R)d\lambda\int_{-\infty}^{\infty}{\rm  e}^{-\lambda\vert\Dz\vert}\jk(\lambda,z')\,dz',
\label{eq:Aphibess}
\end{equation} 
where
\begin{equation}
  \jk(\lambda,z')=\int_0^{\infty}\xi\,\Jo(\lambda \xi) j(\xi,z')\,d\xi,
  \label{eq:hankj}
\end{equation}
is the Hankel transform of order 1 of the current density:
reassuringly, by inverting order of integration in the triple integral
in Equation (\ref{eq:Aphibess}), and performing first the integration
over $\lambda$, eq.~(2.12.38.1) in Prudnikov et al. (1986, hereafter
P86) proves that the resulting expression coincides\footnote{The
  equivalence in the special case of $\Dz=0$, i.e., for points in the
  plane of razor-thin currents, can be also proved by using
  eq.~(2.12.31.1) in P86, and eq.~(6.576.2) in Gradstheyn and Ryzhik
  (2007, hereafter GR07).}  with Equation (\ref{eq:Acylfull}).

From Equation (\ref{eq:rotAcyl}) we then obtain the two components of
the $\Bv$ field
\begin{equation}
\begin{cases}
\displaystyle{
\BR(R,z)={8\pi G\over c^2}\int_0^{\infty}\lambda\Jo(\lambda R)d\lambda\int_{-\infty}^{\infty}{\rm sign}(\Dz){\rm e}^{-\lambda\vert\Dz\vert}\jk(\lambda,z')\,dz',}\cr\cr
\displaystyle{    
\Bz(R,z)={8\pi G\over c^2}\int_0^{\infty}\lambda\Jz(\lambda  R)d\lambda\int_{-\infty}^{\infty}{\rm 
  e}^{-\lambda\vert\Dz\vert}\jk(\lambda,z')\,dz',}
\end{cases}
\label{eq:BzfullH}
\end{equation}
where the expression for $\Bz$ derives from the identity
$d [x \Jo(x)]/dx= x\Jz(x)$. The proof of equivalence of Equations
(\ref{eq:BfullK})-(\ref{eq:BzfullH}) is important but quite laborious,
and requires some comment. For $\BR$ we move $\lambda$ in front of
$\Jo(\lambda R)$ inside the integral over $z'$, we recognize a
derivative with respect to $z'$ of the exponential factor and
integrate by part, we then exchange order of integration and finally
integrate in $\lambda$ by using eq.~(2.12.38.1) in P86. For $\Bz$, the
approach is to move the $\lambda$ in front of $\Jz (\lambda R)$ inside
the Hankel transform $\jk$, use the identity
$\lambda\Jo(\lambda\xi)=-d\Jz(\lambda\xi)/d\xi$ and integrate by parts
over $\xi$. Finally, we invert order of integration and use again
eq.~(2.12.38.1) in P86. With this first approach we proved the
equivalence of Equation (\ref{eq:BzfullH}) with the {\it second}
expressions for $\BR$ and $\Bz$ in Equation (\ref{eq:BfullK}).

However, there is a second approach that proves the equivalence of
Equation (\ref{eq:BzfullH}) with the {\it first} identities in
Equation (\ref{eq:BfullK}), and that also help to clarify an important
convergence issue. In both the triple integrals in Equation
(\ref{eq:BzfullH}) we exchange order of integration, and we evaluate
first the integrals over $\lambda$, that belong to the family
$\int_0^{\infty}x {\rm e}^{-p x}{\rm J}_{\mu}(a x) {\rm J}_{\nu}(b
x)dx$, with the aid of eq.~(2.12.38.2) in P86, where in particular
$p=\vert\Dz\vert$ and $x=\lambda$.  It can be proved (for example by
asymptotic expansion of the Bessel functions for large values of their
argument) that for $\Dz=0$ the two integrals over $\lambda$ diverge:
however, if the limit for $\Dz\to 0$ is evaluated after integration,
then the first identities in Equation (\ref{eq:BfullK}) are
recoverd\footnote{The delicay of the exchange of the limit with the
  integral when using Bessel functions is best illustrated in
  electrodynamics by the case of the magnetic field produced by a
  circular current loop: if one compute the magnetic field in the
  plane of the spire after restricting $\Av$ to the $z=0$ plane, then
  $\Bv$ is predicted to diverge everywhere in the $z=0$
  plane. Instead, if the field is computed for $z\neq 0$, and then the
  limit for $z\to 0$ is considered, the correct expression is obtained
  (see also Exercise 5.10 in J98).} after expressing
$\partial\funU/\partial z'$ and $-\partial\funZ/\partial\xi$ in
explicit form.

\section{Series solution for the gravitomagnetic equations: the
  razor-thin disk with circular orbits}

Having established the general setting of the problem, we are now in
position to discuss the radial dependence of the circular velocity of
a test mass (a star or a gas cloud) in the plane of a razor-thin disk
made of field stars in circular orbits, so that
\begin{equation}
  \rho(R,z)=\Sigma(R)\delta(z),\qquad\vv= v(R)\ephi,\qquad\jv=\chi (R)\delta(z)\ephi, 
\label{eq:rtd}  
\end{equation}
where $\Sigma(R)$ and $v(R)$ are respectively the mass surface
density, and the circular velocity of the disk, and
$\chi(R)=\Sigma(R)\,v(R)$ is the radial profile of the two-dimensional
current; according to the orientation of the coordinate system, a
positive $v$ means counterclockwise rotation.  From the assumption of
circular velocities, it follows necessarily that the modulus of the
circular velocity of the test stars and of the rotating disk coincide,
i.e. $\Vert\vvs(R)\Vert=\vert v(R)\vert$, and quite naturally we also
assume that $\vvs(R)=v(R)\ephi$, i.e. that the test stars rotate as
the stars of the disk. From Equation (\ref{eq:fma}), where the field
$\Ev$ is just the newtonian gravitationl field produced by
$\Sigma(R)$, and from the fact proved in Section 2.1 that in the $z=0$
plane $\Bv=\Bz(R)\ez$, we obtain the scalar equation for the circular
velocity $v$ at each radius $R$
\begin{equation}
v^2=\vO^2 + R v \Bz [\chi],
\label{eq:vcircdisk}
\end{equation}
where $\vO$ is the circular velocity of the disk in the newtonian case
and where, from Equations (\ref{eq:BfullK})-(\ref{eq:BzfullH}), the
gravitomagnetic field in the disk plane reduces to the equivalent
expressions
\begin{equation}
  \Bz [\chi] = {G\over c^2}\times
\begin{cases}
\displaystyle{16\int_0^{\infty}{d\, \xi \chi(\xi)\over d\xi}{\Kc (k)\over R+\xi}\,d\xi,\qquad k = {2\sqrt{R \xi}\over R+\xi},}\cr\cr
\displaystyle{8\pi\int_0^{\infty}\lambda\Jz(\lambda R)\chik(\lambda)d\lambda.}
\end{cases}
\label{eq:Bdisk}  
\end{equation}
The general discussion in Sections 2.2-2.3 shows that the first
expression contains a logarithmic integrable singularity at $\xi =R$,
while in the second the order of integration cannot be exchanged with
the Hankel transform $\chik$ given by Equation (\ref{eq:hankj})
applied to $\chi(R)$.  Notice that with $\Bz [\chi]$ we indicate the
{\it linear} operator acting on the mass current profile (i.e., on the
$v(R)$ profile, if the surface density $\Sigma(R)$ profile is
assigned); notice also that Equation (\ref{eq:vcircdisk}) is a {\it
  nonlinear} integral equation for $v(R)$, even if obtained in the
framework of linearized GR, and it is invariant under the inversion of
the rotation field of the disk, $v(R)\to - v(R)$. In practice, by
solving Equation (\ref{eq:vcircdisk}) for assigned $\Sigma(R)$, we
obtain (at the order of the gravitomagnetic equations) the
``self-consistent'' circular velocity of the disk produced by the
combined effects of the newtonian field and of the gravitomagnetic
potential produced by the rotation curve itself.

\begin{figure*}
\centering 
  \hskip -1truecm 
 \includegraphics[width=0.5\linewidth, keepaspectratio]{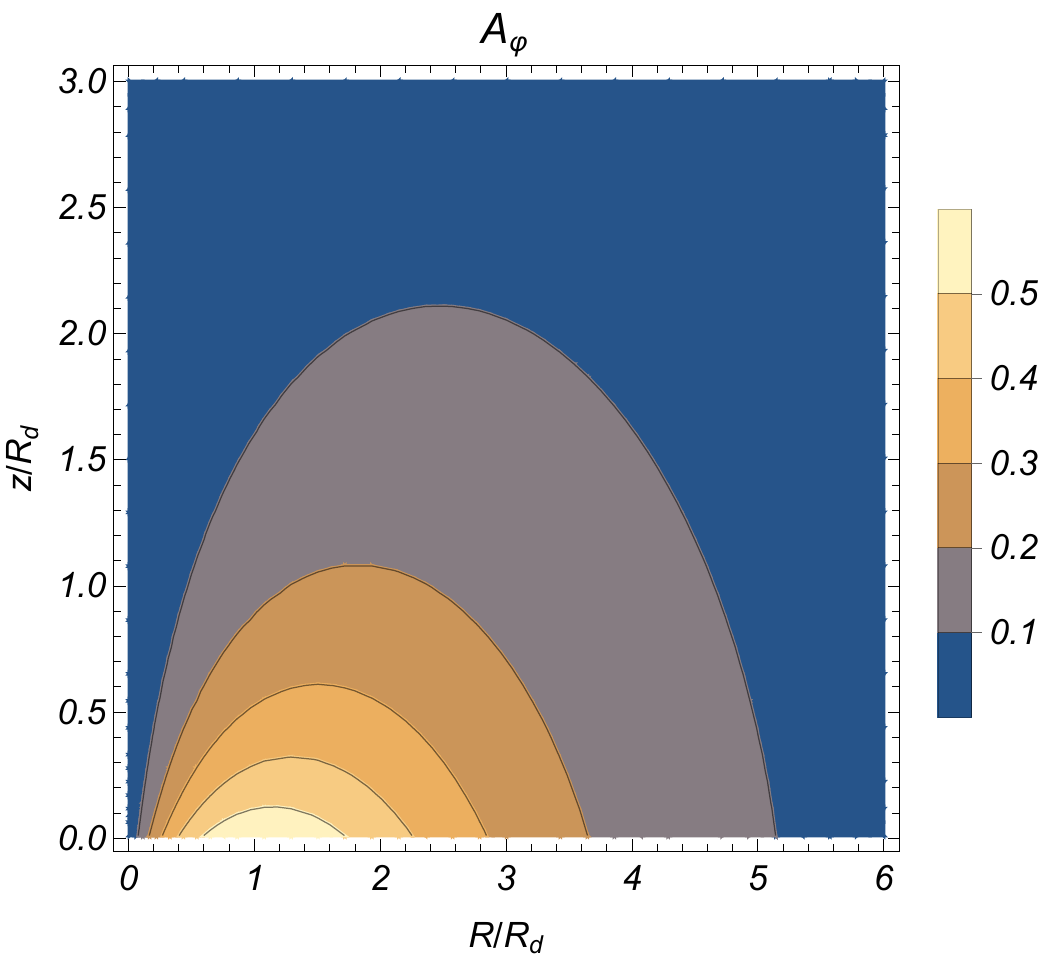}
 \includegraphics[width=0.5\linewidth, keepaspectratio]{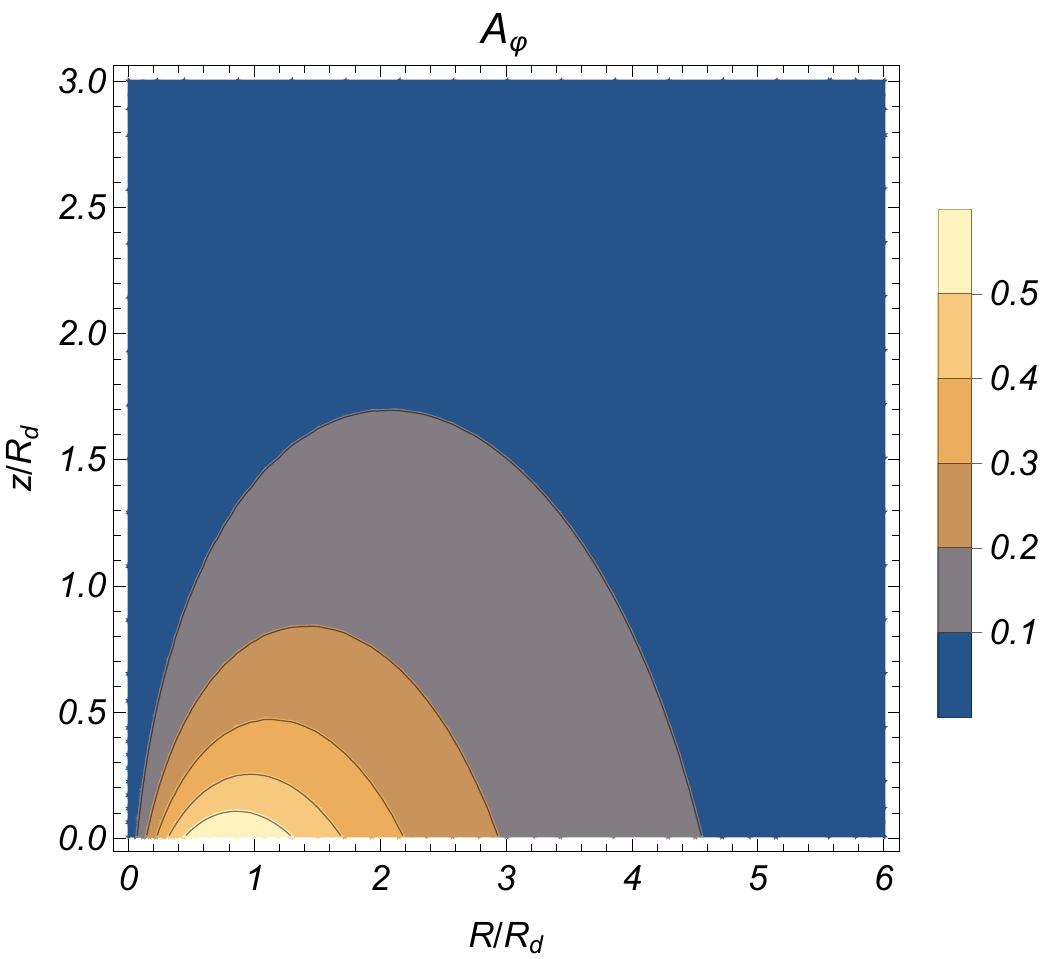}\\
   \hskip -0.5truecm 
\includegraphics[width=0.495\linewidth, keepaspectratio]{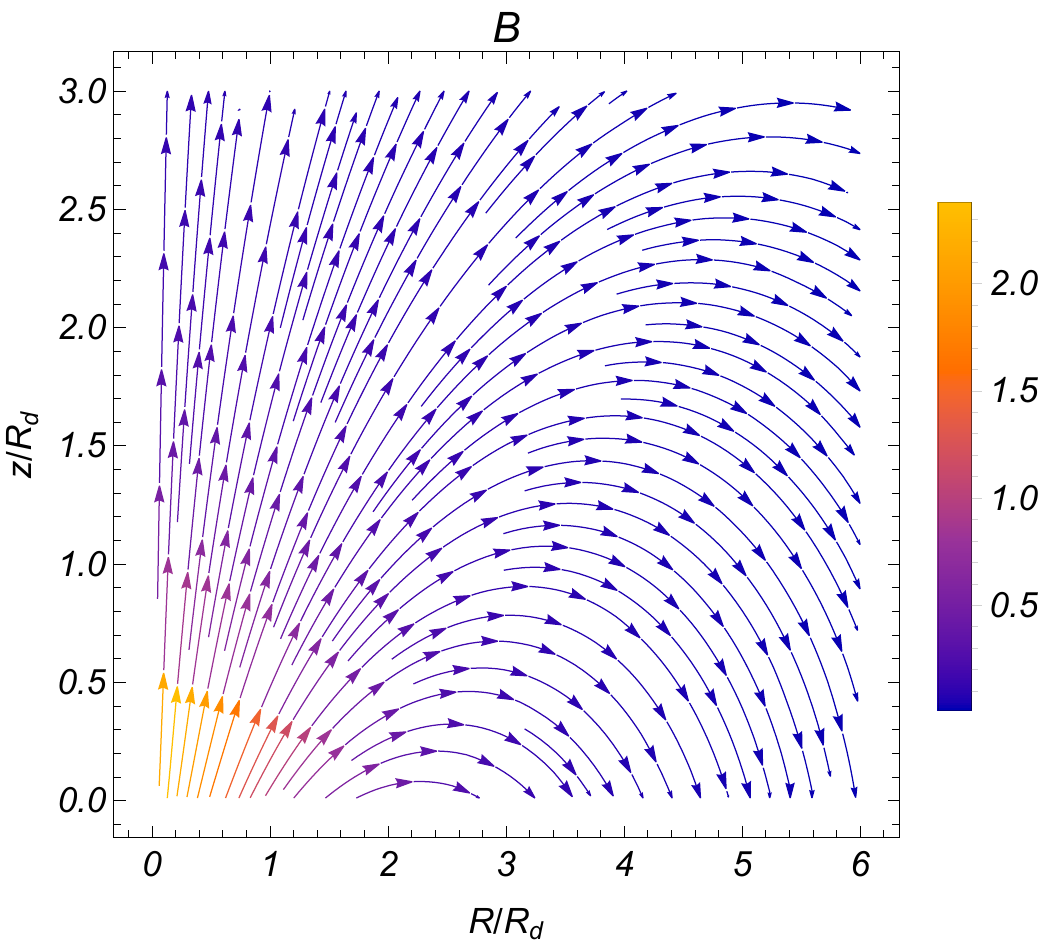}\quad
 \includegraphics[width=0.495\linewidth, keepaspectratio]{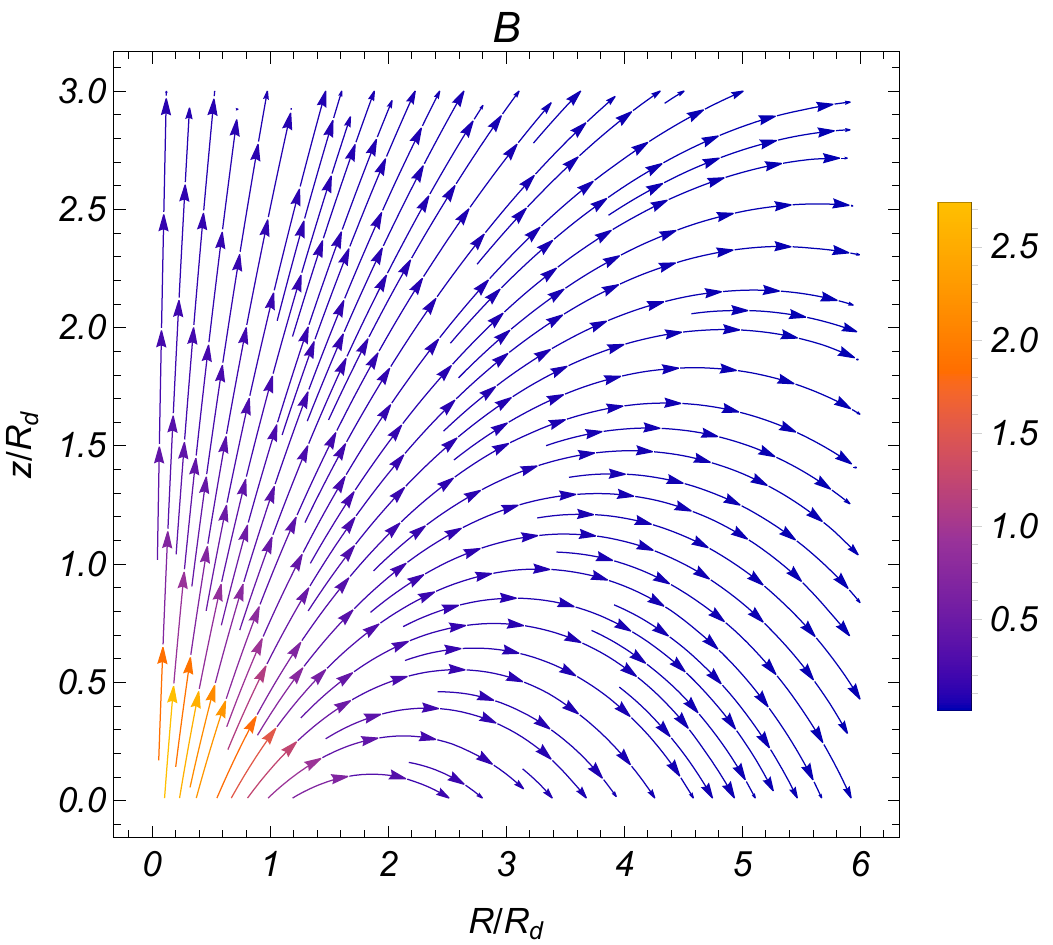}
 \caption{Top panels: the $\Aphi$ field for the counterclockwise
   rotating exponential (left) and Kuzmin (right) razor thin disks, in
   units of $\epsilon \sqrt{G\Md/\Rd}$, where
   $\epsilon = G\Md/(\Rd c^2)$. The field is computed for the
   newtonian current $\chi_0$, as required by Equation
   (\ref{eq:vcdsol}), and it looks similar to the field that would be
   produced by the current of a circular spire; the maximum of the
   newtonian current is located at $R_0\simeq 0.574\Rd$ for the
   exponential disk, and at $R_0\simeq 0.535\Rd$ for the Kuzmin
   disk. The maximum of $\Aphi$ is located at $R_A\simeq 1.12\Rd$ for
   the exponential disk and at $R_A\simeq 0.85\Rd$ for the Kuzmin
   disk.  Bottom panels: the corresponding $\Bv$ field, in units of
   $\epsilon\sqrt{G\Md/\Rd^3}$. At $z=0$, $\Bz$ is positive in the
   inner regions of the disk, and negative outside; the radius $R_B$
   at which $\Bz =0$ is determined by the critical point of
   $R\Aphi(R)$, and so does not coincide with the position of the
   maximum of $\Aphi(R)$, as apparent from the figures. Numerically,
   $R_B\simeq 2.2\Rd$ for the exponential disk, and $R_B\simeq 1.68$
   for the Kuzmin disk.}
  \label{f1}
\end{figure*}
\vskip 0.8truecm 

\begin{figure}
\centering 
  \hskip 0.1truecm 
 \includegraphics[width=0.55\linewidth, keepaspectratio]{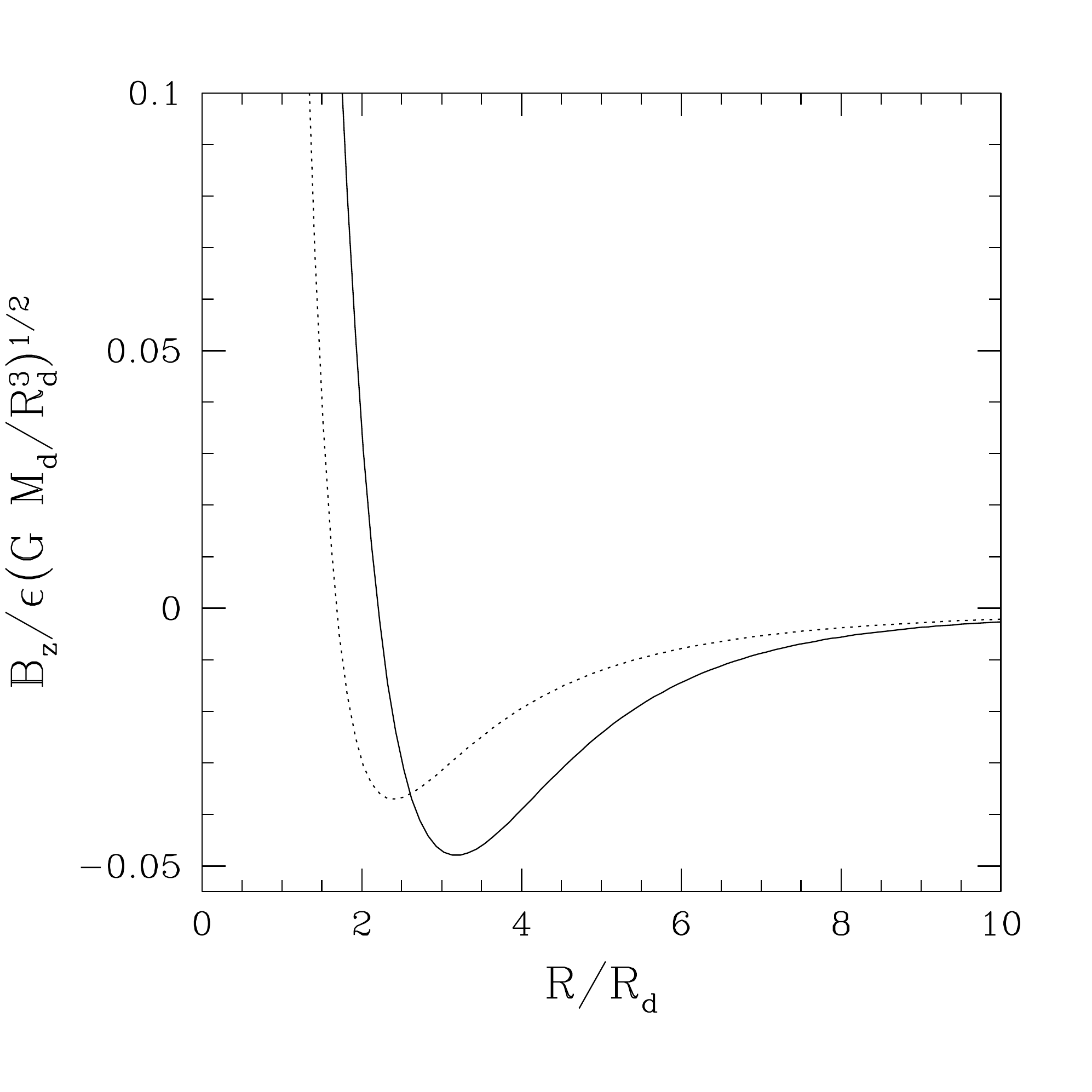}
 \caption{The (dimensionless) gravitomagnetic field $\fun$ in the disk
   plane, for the counterclockwise rotating exponential (solid line),
   and Kuzmin (dotted line) disks made by purely circular orbits,
   computed from the newtonian surface current density $\chi_0$. $\Bz$
   in the central regions of the disks is positive, i.e. directed
   along $\ez$, while in the outer parts $\Bz$ and $\ez$ are
   antiparallel, as also apparent from the bottom panels in Figure
   \ref{f1}.}
  \label{f2}
\end{figure}
\vskip 0.8truecm 

For a razor-thin disk of total mass $\Md$ and scale-lenght $\Rd$, it
is natural to normalize lenghts to $\Rd$, surface densities to
$\Md/\Rd^2$, and velocities to $\sqrt{G\Md/\Rd}$: a tilde over a
quantity indicates normalization to its associated scale.
Accordingly, Equation (\ref{eq:vcircdisk}) is recast in dimensionless
form as
\begin{equation}
{\tilde v}^2=\vOt^2+\epsilon\,\Rtil\,\tilde v\,\fun [\tilde\chi],\qquad \epsilon =
{G\Md\over\Rd c^2},
\label{eq:vcdnor}
\end{equation}
where $\epsilon$ is a dimensionless parameter arising naturally from
the normalization of $\Bv$. In fact, from Equation (\ref{eq:Bdisk}) we
obtain $\Bz = \epsilon \sqrt{G\Md/\Rd^3}\fun$, where $\fun$ is the
dimensionless gravitomagnetic field in the disk.  Notice that for a
disk the parameter $\epsilon$ is nothing else that the natural proxy
for the quantity $v^2/c^2$ mentioned in the Introduction, where $v$ is
a charaterstic velocity associated to the newtonian gravitational
field (see the solid lines in Figure 3). Therefore, in real galaxies
$\epsilon$ is very small: for example $\epsilon\simeq 10^{-6}$, in a
disk galaxy with a characteristic rotational velocity of $300$ km/s.
Accordingly, we represent $v$ as a regular asymptotic series in powers
of $\epsilon$
\begin{equation}
v(\epsilon, R)=\vO (R)+\epsilon^{\alpha}\vo (R)+{\cal
  O}(\epsilon^{2\alpha}),\qquad \alpha>0,
\label{eq:vser}
\end{equation}
where the exponent $\alpha$ is to be fixed by order balance, and of
course for $\epsilon\to 0$ we reobtain the newtonian case. Inserting
the expansion above in Equation
(\ref{eq:vcircdisk}), from the linearity of $\Bz[\chi]$ on velocity,
it follows necessarily that $\alpha =1$, and at points where $\vOt\neq 0$
\begin{equation}
\vot={\Rtil\over 2}\fun [{\tilde\chi}_0], 
\label{eq:vcdsol}
\end{equation}  
where ${\tilde\chi}_0$ is the (normalized) surface density current
profile corresponding to the newtonian rotation curve $\vOt$: from
${\cal O}(\Bz[\tilde\chi_0]) =\epsilon$, it follows that the term
$\vot$ depends on the gravitomagnetic field produced by the newtonian
current only. At the origin, the only point\footnote{A clear
  distinction should be made between the mass current and the rotation
  curve: in case of truncated disks the former is spatially limited to
  the region occupied by the disk, while the latter, and the field
  $\Bz[\chi]$, are defined also in the empty region beyond the disk
  edge.} at finite $R$ where $\vOt$ can vanish, the perturbation
analysis shows that again $\alpha =1$, but
$\vot=\Rtil\fun [{\tilde\chi}_0]$: however, as $R=0$, no difference
arises with Equation (\ref{eq:vcdsol}), that therefore can be used
uniformly over the whole radial range.  Moreover, Equation
(\ref{eq:vcdsol}) shows that the GR corrections due to a
counterclockwise current lead to a {\it decrease} of the rotational
speed where $\Bz[\chi_0]$ is negative, and an {\it increase} where
$\Bz[\chi_0]$ is positive: of course, it is immediate to verify that
the effects on the circular speed are the same for a global inversion
of the rotational velocity $\vO\to -\vO$, due to the associated change
of sign of $\Bz[\chi_0]$.  It is important to remark that Equation
(\ref{eq:vcdsol}) allows for an alternative interpretation: one could
just consider the field $\Bz[\tilde\chi_0]$ at the r.h.s of Equation
(\ref{eq:vcircdisk}), and then solve in closed form the resulting
quadratic equation for $v(R)$. After selecting the sign so that for
$\epsilon\to 0$ the newtonian profile is reobtained, an expansion of
the solution for $\epsilon\to 0$, and truncation at the first order,
gives again Equation (\ref{eq:vcdsol}), proving again the the linear
gravitomagnetic formulation of the problem leads to a regular
perturbation problem (e.g., see Bender and Orszag 1978, Chapter 7).

Of course, even if the problem admits a regular perturbation approach,
with a solution reducing to the newtonian one in the limit
$\epsilon =0$, the function $\vot$ could attain very large (but
finite) values, compensating the small (but non-zero) value of
$\epsilon$, and producing a perturbation of the same order of
magnitude\footnote{In converging regular expansions higher order terms
  can be larger than lower order terms: as a simple example consider
  the first two terms of the absolutely converging series
  ${\rm e}^{\epsilon x}=1 + \epsilon x +{\cal O}(\epsilon^2 x^2)$, for
  fixed $\epsilon$< and $x>\epsilon$.} or even larger than the
newtonian term: this would be a strong support to the possibility of a
GR origin of the flat rotation curve of disk galaxies. In the next
examples, based on realistic disk density profiles, we show however
that this is not the case, and $\vot$ remains small, with absolute
values well below the unity.

\subsection{The exponential disk}

In our first application we consider the razor-thin exponential disk
of total mass $\Md$ and scale-lenght $\Rd$, the standard model used to
describe the stellar density distribution of disk galaxies (e.g.,
BT08, Bertin 2014). The surface density is given by
\begin{equation}
\Sigma(R)={\Md\over\Rd^2} {{\rm e}^{-\Rtil}\over 2\pi}, \qquad \Rtil={R\over\Rd},
\label{eq:Sexpdisk}
\end{equation}
and in newtonian gravity the gravitational potential in the equatorial
plane is
\begin{equation}
\phi (R)=-{G\Md\over\Rd}{\Rtil\over 2}\left[
    \Iz\left({\Rtil\over 2}\right)\Ko\left({\Rtil\over 2}\right)-
    \Io\left({\Rtil\over 2}\right)\Kz\left({\Rtil\over 2}\right) 
\right], 
\label{eq:phiexpdisk}
\end{equation}
so that the associated circular velocity is 
\begin{equation}
  \vO^2(R)={G\Md\over\Rd}{\Rtil^2\over 2}\left[
    \Iz\left({\Rtil\over 2}\right)\Kz\left({\Rtil\over 2}\right)-
    \Io\left({\Rtil\over 2}\right)\Ko\left({\Rtil\over 2}\right) 
\right],
\label{eq:Vexpdisk}
\end{equation}
where ${\rm I}_m$ and ${\rm K}_m$ are the modified Bessel functions of
order $m$ (e.g., BT08, C21). As already remarked in the Introduction,
the maximum of $\vO$ is reached at $R\simeq 2.15\Rd$, and in the
range $1.5 <\Rtil <3$ the curve is almost flat even in absence of DM:
notice that inside $3\Rd$ the disk already contains $\simeq 0.8\Md$.
The newtonian current surface density $\chi_0$ for a disk made by
purely circular orbits is then obtained from Equations
(\ref{eq:Sexpdisk})-(\ref{eq:Vexpdisk}): for a global counterclockwise
rotation, it reaches the maximum at $R_0\simeq 0.574\Rd$.

For the exponential disk it turns out that the most efficient way to
compute the gravitomagnetic field is to use the formulation in terms
of elliptic integrals. In particular, as the current density is a
regular function of $R$, the general discussion about convergence
leads to use Equation (\ref{eq:Acylfull}) for the evaluation of
$\Aphi$, and the first and second integrals in Equation
(\ref{eq:BfullK}) for the evaluation of $\BR$ and $\Bz$, respectively:
of course, in the disk plane the latter expression reduces to the
first case in Equation (\ref{eq:Bdisk}).  Numerically, the integrable
divergence at $\xi=R$ is easily treated by splitting the integral from
the origin to $(1-\eta)R$, and from $(1+\eta)R$ to infinity, and
reducing $\eta$ until acceptable convergence is reached (actually,
thanks to absolute convergence, the two $\eta$'s not need to be the
same).  In the following experiment, convergence was already reached
for $\eta = 10^{-3}$, with stable results at 5 significative digits
according to Mathematica NIntegrate function. Importantly, the
numerical agreement with the alternative formulation in terms of
Bessel functions has been also verified over all the disk.

\begin{figure*}
\centering 
  \hskip -1truecm 
 \includegraphics[width=0.52\linewidth, keepaspectratio]{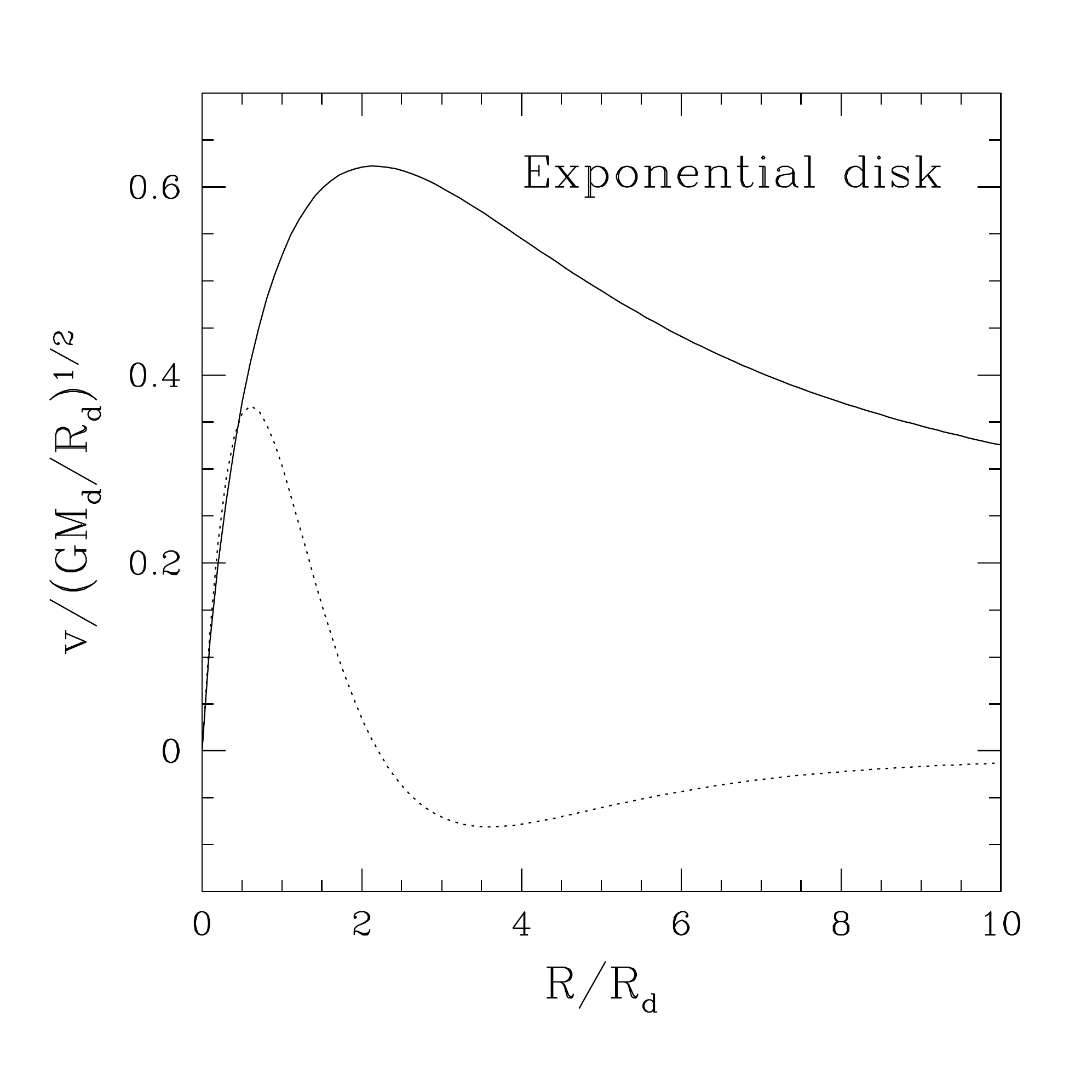}
 \includegraphics[width=0.52\linewidth, keepaspectratio]{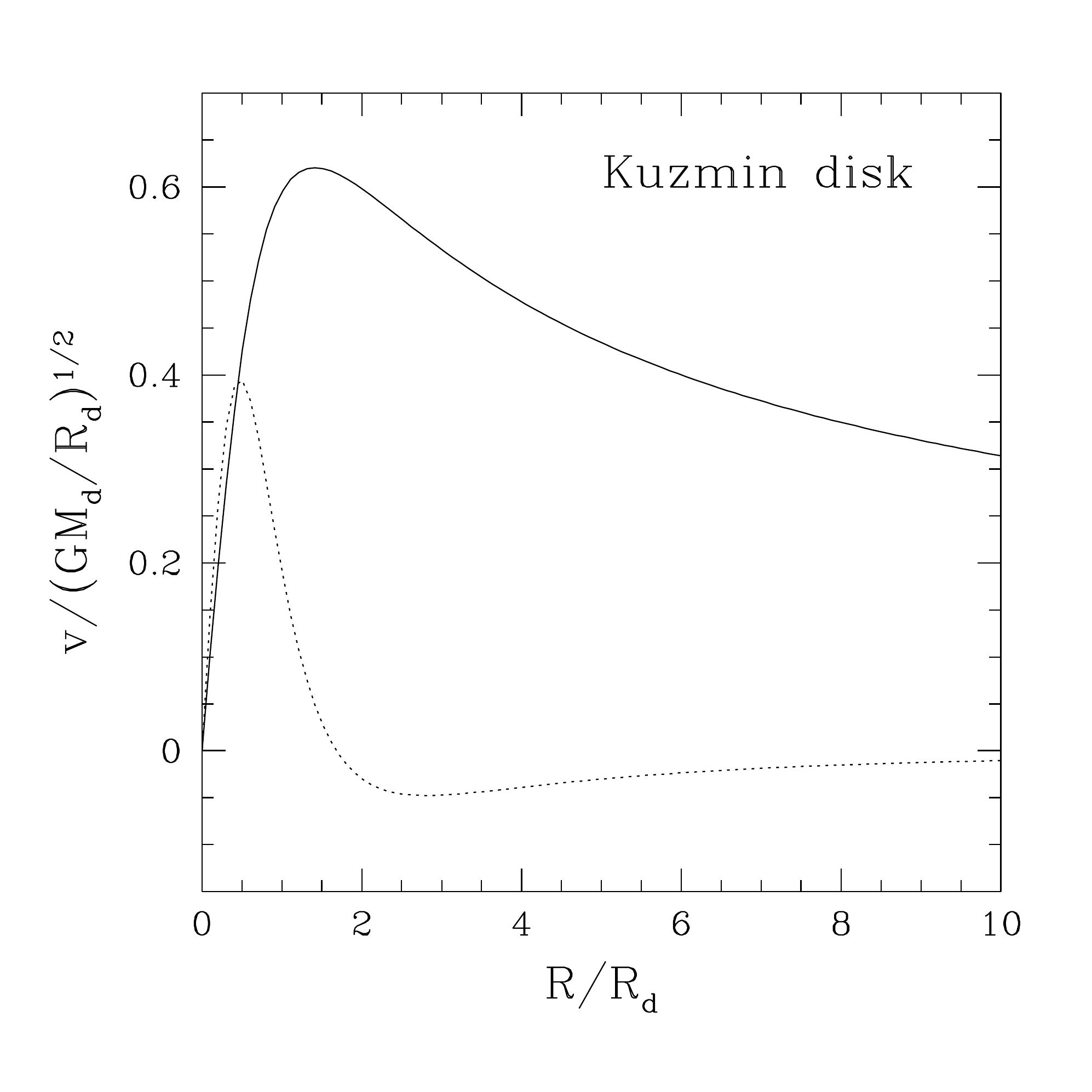}
 \caption{Radial trend of the rotational velocity components in
   Equation (\ref{eq:vser}) normalized to $\sqrt{G\Md/\Rd}$, for the
   exponential (left) and Kuzmin-Toomre (right) disks. The solid line
   is the newtonian rotation curve $\vO$, and the dotted line is the
   $\vo$ perturbative term, that should then be added to $\vO$ after
   multiplication by $\epsilon\simeq 10^{-6}$. Notice how, in
   principle, the effect of the GR gravitomagnetic field at large
   radii is a {\it decrease} of the circular velocity with respect to
   the newtonian case.}
  \label{f3}
\end{figure*}

In the top left panel of Figure \ref{f1}, the only nonvanishing
component $\Aphi$ of the potential vector (in units of
$\epsilon\sqrt{G\Md/\Rd}$), is shown in the meridional plane, where
the overall striking similarity with the field produced by an
``effective'' circular current loop of radius $R_A\simeq 1.12\Rd$ is
apparent, where $R_A$ is the position of the maximum of $\Aphi(R)$;
quite obviously, $R_A$ does not coincide with the position $R_0$ of
the maximum of the current.  In the bottom left panel, the associated
(normalized) $\Bv$ field is shown. As expected, the field is similar
to that of a circular (counterclockwise circulating) current loop,
with $\Bz$ positively directed in the inner regions of the disk (where
the Biot-Savart fields of each current ring composing $\chi_0$
reinforces), and negatively directed in the outer regions, due to the
cooperative (negative) contribution of the current in the inner
regions of the disk, while the counteracting positive contributions of
the current in the outer regions of the disk are less and less
important, due to vanishing of the current at large radii. The radial
trend of $\Bz[\chi_0]$ in the disk plane is represented by the solid
line in Figure \ref{f2}. Notice that the radius at which $\Bz(R)=0$ is
$R_B\simeq 2.2\Rd$, not coincident with $R_A$, as from Equation
(\ref{eq:rotAcyl}) $R_B$ is given by the critical point of
$R\Aphi(R)$, while $R_A$ is the critical point of $\Aphi(R)$. Having
determined the field $\Bz$for the exponential disk, from Equation
(\ref{eq:vcdsol}) we finally compute $\vo$. In the left panel panel of
Figure \ref{f3} the solid line shows the (normalized) profile of the
newtonian rotation curve $\vO$, while the dotted line $\vo$. After
multiplication by $\epsilon\simeq 10^{-6}$, it is clear the the
effects of GR on the curve of the exponential disk are well below the
possibility of any practical detection, with differences with respect
to the newtonian rotation curve well below 1 m/s: at the description
level of gravitomagnetism, GR does not produce any effect on the
rotation curve of the disk. In any case, it is interesting to notice
that for the reasons explained above, the sign of the gravitomagnetic
field in the Lorentz equation actually produces a {\it decrease} of
the rotational speed at large galactocentric distances!

\vskip 0.8truecm 

\subsection{The Kuzmin-Toomre disk}

Having determined the solution for the exponential disk, in order
  to confirm the obtained results we move to explore a different disk
  model. We focus on the Kuzmin-Toomre razor-thin disk (Kuzmin 1956,
Toomre 1963, see also BT08, C21), an idealized model widely used
stellar dynamics for its mathematical simplicity. The surface
density-potential pair of the disk of total mass $\Md$ and
scale-lenght $\Rd$ is given by
\begin{equation}
  \Sigma(R) = {\Md\over\Rd^2}{1\over 2\pi (\Rtil^2+1)^{3/2}},\qquad
  \phi(R)=-{G\Md\over\Rd} {1\over \sqrt{\Rtil^2+1}},\qquad \Rtil={R\over\Rd},
 \label{eq:kuzdisk}
\end{equation}  
where the potential is restricted to the disk plane, and the newtonian
circular velocity and the surface mass density current are given by
\begin{equation}
  \vO^2(R)={G\Md\over\Rd} {\Rtil ^2\over (\Rtil^2+1)^{3/2}},\qquad
  \chi_0(R)=\sqrt{G\Md^3\over\Rd^5}{\Rtil\over 2\pi (\Rtil^2+1)^{9/4}}.
\label{eq:Vkuzdisk}
\end{equation}
The maximum of $\vO$ is located at $R=\sqrt{2}\Rd$, and that of
$\chi_0$ at $R_0 =\sqrt{2/7}\Rd\simeq 0.535\Rd$, a value curiously
similar to that of the exponential disk.  We now
show that $\vo$ can be obtained in closed form by using the
Fourier-Bessel identity in Equation (\ref{eq:Bdisk}). In fact, from
eq. (6.565.4) of GR07, or eq. (2.12.4.28) of P86, we obtain the Hankel
transform of order 1 for the newtonian mass current as
\begin{equation}
\chik(\tilde\lambda)=\sqrt{G\Md^3\over\Rd}\,
{{\tilde\lambda}^{5/4}{\rm K}_{1/4}(\tilde\lambda)\over
  2^{9/4}\pi\Gamma (9/4)},\qquad \tilde\lambda=\lambda\Rd,
\label{eq:chi1H}
\end{equation}
where $\Gamma$ is the complete gamma function, and we used the fact
that for modified Bessel functions ${\rm K}_{-\nu}={\rm K}_{\nu}$;
notice that $\tilde\lambda$ is a dimensionless quantity. The last
integral in Equation (\ref{eq:Bdisk}) can be expressed analytically
thanks to eq.~(6.576.3) of GR07 or eq. (2.16.21.1) of P86, and finally
from Equation (\ref{eq:vcdsol}) we obtain
\begin{equation}
  \vo (R)=\sqrt{G\Md\over\Rd}\,
  {2\Gamma(7/4)\Gamma(3/2)\over\Gamma (9/4)}{_2F_1}\left({7\over
      4},{3\over 2};1;-\Rtil^2\right)\,\Rtil,
\end{equation}
where $_2F_1$ is the standard hypergeometric function. Notice that
equation above can be recast in terms of elliptic integrals, resulting
in perfect agreement over the whole radial range with the values of
$\vo(R)$ obtained from Equation (\ref{eq:vcdsol}) and numerical
integration of the last expression of $\Bz$ in Equation
(\ref{eq:BfullK}), in a reassuring validity check. As a mathematical
curiosity, we also notice that for the newtonian current $\chi_0$ of
the Kuzmin-Toomre disk, not only the field $\Bz$ in the disk plane can
be obtained explicitely, but also Equations (\ref{eq:BzfullH}) can be
solved analytically over all the space by using Fox {\rm H} function,s
after expressing Equation (\ref{eq:chi1H}) in terms of these functions
(e.g., Mathai et al. 2010, and in particular eq. (2.25.3.2) in
Prudnikov et al. 1990): the resulting expression is however of no
practical use, and so not reported here.

In the right panels of Figure \ref{f1} we plot the normalized
potential vector of the disk in the meridional plane, and the
associated gravitomagnetic field: the maximum of the gravitomagnetic
potential vector is reached at $R_A\simeq 0.85\Rd$. The similarity
with the case of the exponential disk, and also the range of values
spanned by the normalized fields is remarkably similar, even if the
radial density profiles of the two disks are quite different,
especially at large radii; notice also the similar behavior of $\Bz$
in Figure \ref{f2}, with positive values in the inner regions of the
disk (i.e., for $R<R_B\simeq 1.68$), and negative values outside.  In
Figure \ref{f3} we plot the normalized profiles of $\vO$ and $\vo$. It
is again apparent how $\vo$ remains limited over all the radial range,
confirming that the contribute to the rotational velocity of the first
order perturbative term in the velocity expansion is fully controlled
by the smallness of $\epsilon$, and again the expected GR corrections
to the newtonian rotational velocity of the galaxy are well below the
detection limit, with differences between the newtonian and the
gravitomagnetic GR curve well below 1 m/s. Finally, in accordance with
the change of sign of $\Bz$ along the equatorial plane, the
gravitomagnetic effects produce a decrease\footnote{ The change of
  sign of $\Bz(R)$ is not a universal property: for example, it is
  quite easy to prove that in razor-thin power-law disks (and so of
  infinite total mass), with a uniform sense of rotation, the
  gravitomagnetic field in the disk cannot change sign with $R$.}  of
the rotational speed at large distances from the galaxy center.

Therefore, from the analysis conducted so far, it seems a quite robust
conclusion that GR, at the level of the gravitomagnetic weak field
approximation, does not affect the rotation curve of self-gravitating
barionic disks with realistic density profiles, and made mostly by
circular orbits. In the next Section we relax the assumption of a
two-dimensional density distribution, and we also consider
non-circular orbits for the stars producing the field in the system.

\section{The gravitomagnetic Jeans equations}

After the discussion of razor-thin disks, we address the problem of
the expected effects of GR on the internal dynamics of genuinely
tridimensional (collisionless) astronomical systems, such as
finite-thickness disk galaxies, or axisymmetric elliptical
galaxies. We first rigorously derive the gravitomagnetic modification
of the Jeans equations, and then we restrict to stationary
axisymmetric systems.  As usual, we indicate with $\vv$ the
phase-space velocity, so that the density $\rho$ and the streaming
velocity field $\vbar$ of the system are given by
\begin{equation}
\rho(\xv)=\int f\dtvv,\qquad\vbar (\xv)\equiv\overline{\vv}=
{1\over\rho(\xv)}\int f \vv\dtvv,
\label{eq:moments01}
\end{equation}
where $f(\xv,\vv,t)$ is the phase-space distribution function
(hereafter DF, see BT08, Bertin 2014, C21), and a bar over a quantity
indicates its average over the velocity space. In the following we
will write $\vv={\rm v}_i{\bf e}_i= \vpR\eR+ \vpphi\ephi+ \vpz\ez$,
and $\vbar=\vi {\bf e}_i= \vR\eR+ \vphi\ephi+ \vz\ez$, where sums over
repeated indices hold in Cartesian coordinates.

Suppose the considered stellar system is self-consistent, i.e., the
motion of each star is determined by the field produced by the
combined effects of all the other stars.  As we are in the low
velocity limit, when the Biot-Savart law can be interpreted as the sum
of the Lorentz fields produced by each moving charge (an
interpretation in general {\it not} true for currents produced by
particles of arbitrary large velocities, see e.g. J98,
Chapter 5; Griffiths 1999, Chapter 5; Feynman 1977, Chapters 13 and
21; Panofsky and Phillips 1962, Chapter 7), it follows that the
acceleration experienced by a star at $\xv$ with velocity $\vv$ can be
written by summing the r.h.s. member of Equation (\ref{eq:fma}) over
the DF,
\begin{equation}
{d^2\xv\over dt^2}=-\nabla\phi - \vv\wedge\Bv [\jv],\qquad
\jv(\xv)=\rho\vbar=\int f \vv\dtvv,
\label{eq:lorentzJ}
\end{equation}
where $\phi(\xv)$ is the newtonian gravitational potential of the
system and $\Bv[\jv]$, by virtue of the linearity of $\Bv$ on the
current, is the gravitomagnetic field produced at $\xv$ by the total
streaming current of the system. The hierarchy of the Jeans equations
of increasing order is then obtained by taking moments over the
velocity space of the collisionless Boltzamnn equation (e.g., see
BT08, C21)
\begin{equation}
{\partial f\over\partial t}+\vpi {\partial f\over\partial x_i} +
{d\vpi\over dt}{\partial f\over\partial\vpi}=0,
\label{eq:CBE}
\end{equation}
here written in Cartesian coordinates.  After multiplication of
Equation (\ref{eq:CBE}) by $1$, $\vpi$, $\vpi\vpj$, etc., and
integration, the three second order moments equations
(the gravitomagnetic modification of the usual Jeans equations of
stellar dynamics) are given by
\begin{equation}
{\partial\rho\vi\over\partial t}+{\partial\rho\overline{\vpi\vpk}\over\partial x_k}=
-\rho{\partial\phi\over\partial  x_i}-\epsilon_{ijk}\rho\vj B_k[\jv],\qquad i=1,2,3,
\label{eq:JeansCart}
\end{equation}
where 
\begin{equation}
\overline{\vpi\vpj}(\xv)= {1\over\rho(\xv)}\int f\,\vpi\vpj\dtvv,\qquad
\sigma^2_{ij}(\xv)=\overline{(\vpi -\vi)(\vpj
  -\vj)}=\overline{\vpi\vpj} - \vi\vj;
\label{eq:moments}
\end{equation}
in particular, $\sigma_{ij}^2$ are the components of the second-order
velocity dispersion tensor. In Appendix B, the general gravitomagnetic
time-dependent Jeans equations are also reported in cylindrical
coordinates.

We now to restrict to the axysimmetric stationary case. In the usual
stellar dynamical case, from the Jeans theorem one would assume a
phase-space DF depending on the two classical integrals of the motion
(per unit mass) $E$ and $\Jzero$, i.e. the orbital energy, and the
$z$-component of the angular momentum of stellar orbits. As well
known, under this assumption (e.g., BT08, C21), important constraints
on the allowed velocity moments follow, namely 1) the only possible
streaming motions are in the azimuthal direction, i.e.,
$\vbar=\vphi(R,z)\ephi$, and 2) the velocity dispersion tensor is in
diagonal form at each point in the system, with
$\sigma_R^2=\sigma_z^2\equiv\sigma^2$, while only
$\overline {v_{\varphi}^2}=\sigma_{\varphi}^2+\vphi^2$ remains
determined; when $\sigma_{\varphi}^2=\sigma^2$ everywhere, the system
is said isotropic.  Therefore, for stationary stellar dynamical
two-integral axisymmetric systems with $f(E,\Jzero)$, the current
would be of the ``circular type'' considered in Equation
(\ref{eq:current}), with $\jv=j\ephi=\rho\vphi\ephi$: notice that,
even if the current associated with the streaming velocity field is
circular, the orbits of the individual stars contributing to the
resulting gravitomagnetic field are in general {\it not} circular, and
so we are considering an orbital structure more complicated than that
of the razor-thin disk in Section 3.

However, in the case of stationary collisionless stellar systems with
gravitomagnetic forces, the situation is not so simple, because while
$E$ is still an integral of motion, $\Jzero$ is not conserved along
orbits and so it cannot be used as an isolating integral in the
DF. Fortunately, the problem is solved as follows. As well known, the
lagrangian (per unit mass) associated with Equation
(\ref{eq:lorentzJ}) is
\begin{equation}
 {\cal L}={\Vert\vv\Vert^2\over 2}-(\phi + \Av\cdot\vv),
 \label{eq:lagrang}
\end{equation}
where at variance with the EM case (e.g., see Landau and Lifshitz
1971) the + sign in the generalized potential derives from the - sign
in the gravitomagnetic Lorentz force. Suppose now that
$\vbar=\vphi(R,z)\ephi$, so that $\Av=\Aphi(R,z)\ephi$ from the
results in Section 2.1. The Euler-Lagrange equations show immediately
that $\Jzero$ is not conserved, but a second integral of motion
exists, $I_2=\Jzero -R\Aphi$. The Jeans theorem then in principle
allows for a two-integrals DF $f(E,I_2)$, and the interesting question
arises wether such a DF is consistent with the assumption of streaming
motions with $\vbar=\vphi(R,z)\ephi$ used to prove the existence of
$I_2$.  The answer is in the affirmative, and it is easy to prove that
the properties (1) and (2) mentioned above for systems supported by a
$f(E,\Jzero)$ are preserved by the generalized $f(E,I_2)$, in
particular the fact that $\vR=\vz=0$. Therefore, the vertical and
radial Jeans equations (\ref{eq:jeansCF}) for a stationary,
axisymmetric two-integrals system with gravitomagnetic forces,
supported by a DF of the family $f(E,I_2)$, become
\begin{equation}
  \begin{cases}
\displaystyle{{\partial\rho\sigma^2\over\partial z}=-\rho{\partial\phi\over\partial z}+j\BR[j],\qquad j=\rho\vphi,}\cr\cr  
\displaystyle{{\partial\rho\sigma^2\over\partial R} - {\rho\Delta\over R} =-\rho{\partial\phi\over\partial R} -j\Bz[j],}
\end{cases}  
\label{eq:jeansEJ}
\end{equation} 
where $\Delta=\overline{\vpphi^2}-\sigma^2$, and in the isotropic
rotator case $\Delta =\vphi^2$. Finally, the azimuthal Jeans equation
vanishes identically also in the gravitomagnetic case, as obvious from
the last of Equations (\ref{eq:jeansCF}).  Notice how at variance with
the classical newtonian case, now the vertical and radial velocity
dispersions depend on the ordered rotational field $\vphi(R,z)$, yet
to be determined.

Equations (\ref{eq:jeansEJ}) can be formally solved in terms of
$\sigma^2$ and $\Delta$ as in the usual stellar dynamical case,
obtaining
\begin{equation}
\rho\sigma^2=\int_z^{\infty}\rho{\partial\phi\over\partial z'}\,dz' - U  = \rho\sigma_0^2 -U,\qquad
U(R,z) =\int_z^{\infty}j\BR[j]\, dz', 
\label{eq:solj1}
\end{equation}
where $\sigma_0^2$ is the newtonian solution, and $U(R,z)$ accounts for
the gravitomagnetic effects. Moreover, some algebra shows that
\begin{equation}
{\rho\Delta\over R} = C[\rho,\phi]  - V,\qquad
C[\rho,\phi]=\int_z^{\infty}\left({\partial\rho\over\partial
    R}{\partial\phi\over\partial z'} -{\partial\rho\over\partial z'}{\partial\phi\over\partial R}\right)\,dz'= {\rho\Delta_0\over R}
\label{eq:solj2}
\end{equation}
where the commutator $C[\rho,\phi]$ is the standard newtonian solution
(e.g., Hunter 1977, Barnab\`e et al. 2006, see also C21), while the
term
\begin{eqnarray}
  V(R,z)&=&\int_z^{\infty}\left({\partial j\BR\over\partial R} + {\partial j\Bz\over\partial z'}\right)\,dz' = 
                 \int_z^{\infty}\left[\left({\partial\over\partial R}{j\over R}\right)R\BR+  {\partial j\over\partial z'}\Bz\right]\,dz'\cr  
         &=& - C[j,\Aphi] -  {j\Aphi\over R},
\label{eq:solj3}
\end{eqnarray}
contains the gravitomagnetic effects. The triple integral in the
second expression above (obtained from the first by exploiting the
solenoidal nature of $\Bv$) is particularly well suited for numerical
computations when the current is analitically available (see next
Section), while the last expression is obtained from the first taking
into account that $\Bv=\nabla\wedge\Av$, Equation (\ref{eq:rotAcyl}),
and finally integrating by parts with $j\Aphi =0$ for $z\to\infty$.

As noticed, the difficulty in the solution of the gravitomagnetic
Jeans equations is the fact that the force field depends on the
azimuthal streaming velocity distribution, a quantity that it is not
known in advance. In principle, to solve the gravitomagnetic Jeans
equations one should 1) impose a decomposition on the still unknown
azimuthal velocity field, for example by using the widely adopted
Satoh (1980) ansatz $\vphi^2=k^2\Delta$ (e.g., see C21), 2) solve the
resulting nonlinear integrodifferential Equation (\ref{eq:solj2}) for
$\vphi$ and obtain $\BR[j]$, 3) integrate Equation (\ref{eq:solj1})
and obtain $\sigma^2$, 4) conclude by determining $\Delta$ and the
tangential velocity dispersion
$\sigma_{\varphi}^2=\sigma^2+(1-k^2)\Delta=
\sigma^2+(1/k^2-1)\vphi^2$. Quite obviously, a considerable
simplification is achieved by exploiting the fact that in Equation
(\ref{eq:jeansEJ}) the fields $\BR$ and $\Bz$ are proportional to the
small parameter $\epsilon = G\Mstar/(\rstar c^2)$, where $\Mstar$ is
the total mass of the system, and $\rstar$ its scale-lenght. As in the
razor-thin disks, it follows immediately that the first order
perturbative terms in the natural expansions
$\sigma^2=\sigma_0^2 +\epsilon\sigma_1^2 \ldots\;$,
$\Delta=\Delta_0 +\epsilon\Delta_1 \ldots\;$, and
$\vphi=v_{\varphi 0}+\epsilon v_{\varphi 1} \ldots\;$,
are related to the functions $U$ and $V$ evaluated
over the newtonian current $j_0=\rho \vphiz$, where $\vphiz (R,z)$ is
the streaming rotation velocity obtained from the classical Jeans
equations.

We are now in position to evaluate the effects of GR corrections on
the circular velocity $v(\epsilon,R)=\vO(R)+\epsilon\vo(R) +\ldots\;$
of a test star in the equatorial plane of an axisymmetric
collisionless system described by Equations (\ref{eq:jeansEJ}). The
only difference with the razor-thin disks in Section 3 is that now the
current is provided by the azimuthal streaming velocity field of the
system. From Equation (\ref{eq:lorentzJ}), the analogous of Equations
(\ref{eq:vcircdisk}) and (\ref{eq:vcdnor}) still apply, and
\begin{equation}
\vot={\Rtil\over 2}\fun [{\tilde\j}_0], 
\label{eq:vcirMN1}
\end{equation}
$\Bz = \epsilon \sqrt{G\Mstar/\rstar^3}\fun$ is computed in the disk
plane, and finally $j_0=\sqrt{G\Mstar^3/\rstar^7}\,{\tilde\j}_0$.

The expression for the first order correction term to the streaming
velocity (restricting for simplicity to the isotropic rotator case for
the stellar distribution of the galaxy) is instead obtained from
Equation (\ref{eq:solj2}), considering that at the lowest order
$V ={\cal O}(\epsilon)$, and so from
$\vphi=v_{\varphi 0}+\epsilon v_{\varphi 1} \ldots\;$, after order
balance and some simplification, one gets
\begin{equation}
\tilde v_{\varphi 1} = - {\Rtil\over 2\tilde\j_0}{\cal V}[{\tilde\j}_0], 
\label{eq:vphiMN1}
\end{equation}
where $V= \epsilon (G\Mstar^2/\rstar^5) {\cal V}$ is computed from
Equation (\ref{eq:solj3}).

\subsection{The Miyamoto-Nagai disk}

We now solve the gravitomagnetic Jeans equations, and evaluate
Equations (\ref{eq:vcirMN1})-(\ref{eq:vcirMN1}) for the widely used
Miyamoto-Nagai (1975, Nagai and Miyamoto 1976, see also BT08, C21)
disk of total mass $\Mstar$ and scale-lengths $a$ and $b$, with
newtonian density-potential pair
\begin{equation}
 \rho(R,z) = {\Mstar\over a^3} {s^2\over 4\pi}
 {\Rtil^2+(1+3\zeta)(1+\zeta)^2\over \zeta^3\,[\Rtil^2 +(1+\zeta)^2]^{5/2}},\qquad 
 \phi(R,z)= -{G\Mstar\over a}{1\over\sqrt{\Rtil^2+(1+\zeta)^2}}, 
\label{eq:MNdisk}
\end{equation}
where $\zeta = \sqrt{\ztil^2+s^2}$, $\Rtil=R/a$, $\ztil=z/a$, and the
newtonian circular velocity in the equatorial plane is given by
\begin{equation}
\vO^2(R)={G\Mstar\over a} {\Rtil^2\over[\Rtil^2+(1+s)^2]^{3/2}}. 
\label{eq:MNvcirc0}
\end{equation}
the dimensionless parameter $s=b/a$ measures the disk flattening, and
for $b=0$ the model reduces to the Kuzmin disk of total mass $\Mstar$
and scale-lenght $a$, so we are in the condition to explore the GR
effects of a non-zero thickness on the rotation curve, by comparison
with the case in Section 3.2.

We restrict for simplicity to the isotropic rotator, with a
conterclockwise streaming velocity field, so that the newtonian
azimuthal streaming velocity field and the associated current density
are
\begin{equation}
\begin{cases}
\displaystyle{
\vphiz^2(R,z) ={G\Mstar\over a} {\Rtil^2\over [\Rtil^2 + (1+3\zeta)(1+\zeta)^2]\sqrt{\Rtil^2+(1+\zeta)^2}},
}\cr\cr 
\displaystyle{
j_0(R,z)=\sqrt{{G\Mstar^3\over a^7}}{s^2\over 4\pi}
{\Rtil \sqrt{\Rtil^2 + (1+3\zeta)(1+\zeta)^2}\over\zeta^3 [\Rtil^2+(1+\zeta)^2]^{11/4}},
}
\end{cases}
\label{eq:MNcurr}
\end{equation}
(e.g., Ciotti and Pellegrini 1996, Smet et al. 2015). Notice that in
the equatorial plane the location of the maximum of $j_0(R,0)$ can be
evaluated analytically as a biquadratic equation, and for $s\to 0$ the
position coincides with that of the maximum for the Kuzmin disk.

\begin{figure*}
\centering 
  \hskip -1truecm 
 \includegraphics[width=0.52\linewidth, keepaspectratio]{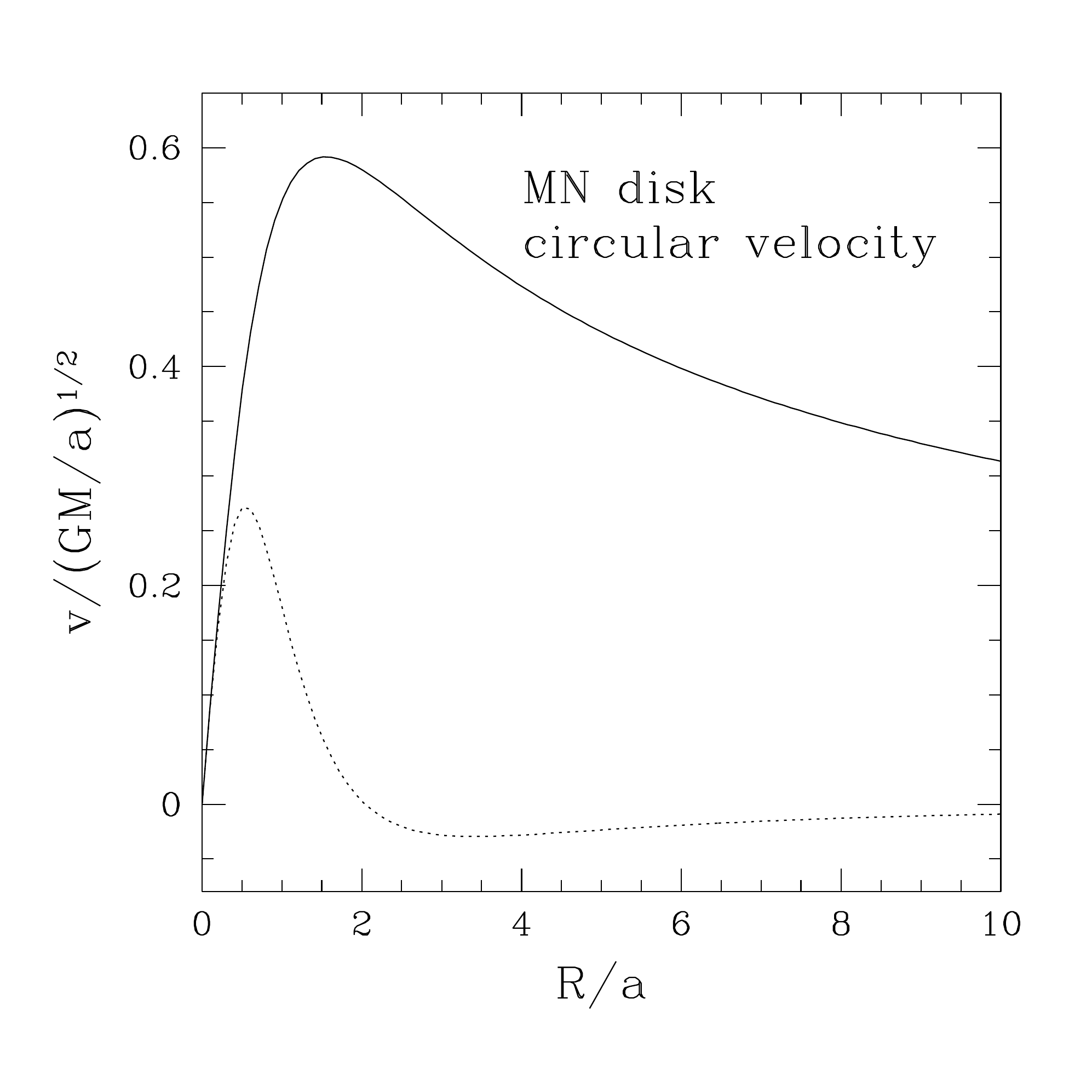}
 \includegraphics[width=0.52\linewidth, keepaspectratio]{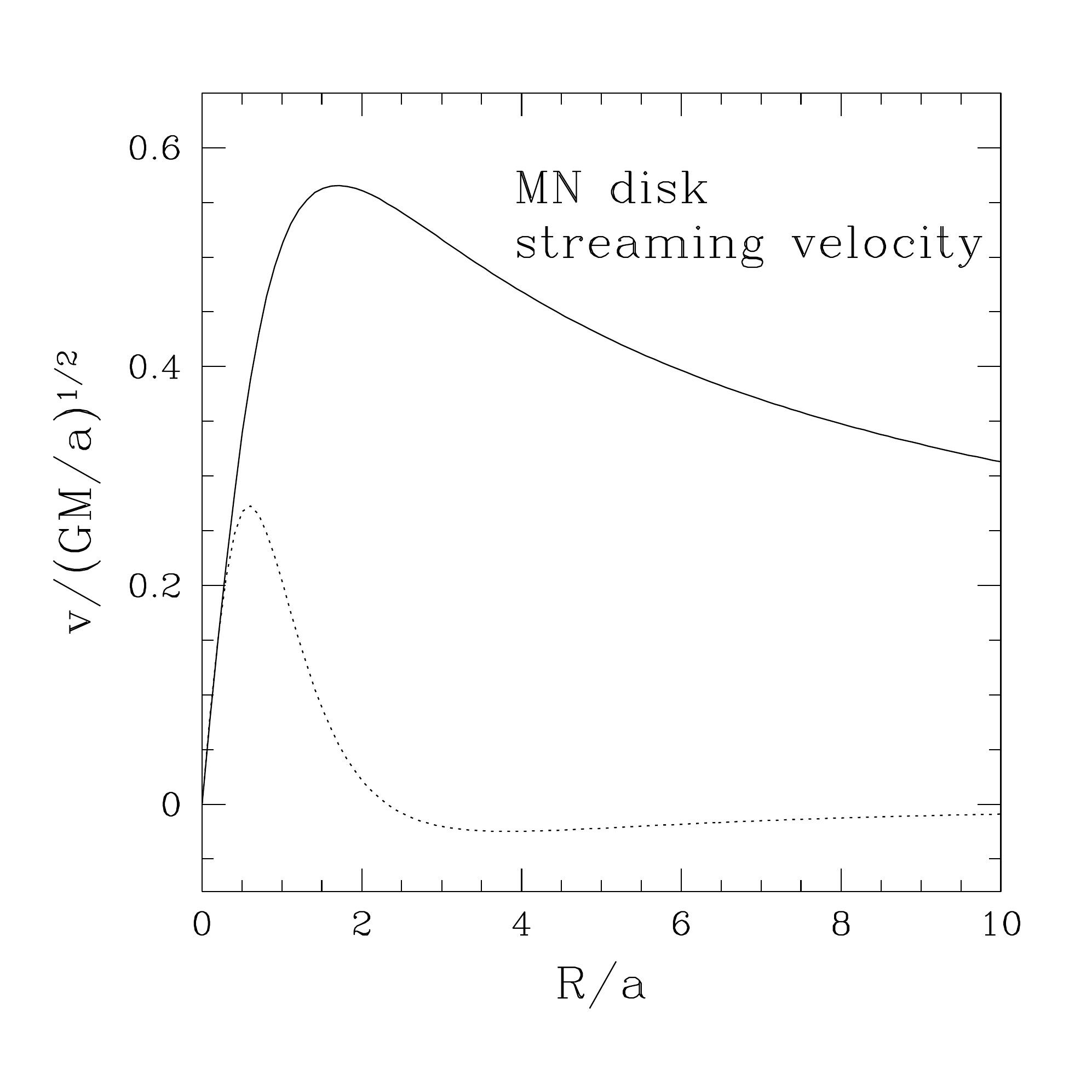}
 \caption{Radial trend of the rotational velocity components (left),
   and of the streaming velocity (right), normalized to
   $\sqrt{G\Mstar/a}$, in the equatorial plane of the counterclockwise
   rotating isotropic Miyamoto-Nagai disk with flattening parameter
   $s=b/a=0.1$. The solid lines are the newtonian components, and the
   dotted lines are the first perturbative term, that should be added
   to the corresponding newtonian term after multiplication by
   $\epsilon\simeq 10^{-6}$. Note how the circular velocity is
   slightly larger than the streaming velocity, the well known
   phenomenon of asymmetric drift produced by the disk vertical
   velocity dispersion support.}
  \label{f4}
\end{figure*}

In the left panel of Figure \ref{f4} we show the results for the
circular velocity, to be compared with the analogous Figure \ref{f3}
for the two razor thin disks made by circular orbits. The solid line
represents the newtonian circular velocity $\tilde v_0$ in Equation
(\ref{eq:MNvcirc0}), while the dotted line gives the correction term
in Equation (\ref{eq:vcirMN1}), where the gravitomagnetic field is now
produced by stars that in general are not moving in circular
orbits. Again, for realistic mass distributions, the perturbation term
shoud be multiplied by factor $\epsilon\approx 10^{-6}$ before
addition to the newtonian term: the GR effects appear again to be
completely negligible and, in analogy with the case of razor thin
disks, positive in the inner regions of the galaxy, and negative at
large radii, {\it reducing} there the rotational speed of a test
mass. In the right panel we show instead the streaming velocity of
stars in the disk equatorial plane, both the newtonian component
(solid line), and the lowest order GR term (dotted line). Again, it is
important to remark that the orbits of the individual stars producing
the $\vphi$ in the equatorial plane are in general {\it not} circular,
and not restricted to the equatorial plane. Notice how the newtonian
streaming velocity is slightly below the newtonian circular velocity,
the well known phenomenon of {\it asymmetric drift}, due to effect of
the vertical component of the velocity dispersion tensor: (e.g., see
BT08). Notice also how the asymmetric drift (a small but well measured
phenomenon in real galaxies) is several order of magnitude larger than
the GR corrections.

The results for the Miyamoto-Nagai disk reinforces the previous
conclusions, i.e., also in three dimensional galaxies, with stars
moving on non-circular orbits, the GR effects predicted by linear
gravitomagnetism on the rotational velocities are a factor of
$\approx 10^{-6}$ smaller than the newtonian predictions. If newtonian
gravity requires DM, unless ``strong'' weak-field GR effects (beyond
the linear gravitomagnetic description) actually dominates the
dynamics of the galaxies, exactly the same amount and distribution of
DM is required in GR and in newtonian gravity, in order to reproduce
the observed rotational curves. 

\vskip 0.8truecm 

\section{Discussion and conclusions}

Recently, it has been suggested that the observed flat radial profile
of the rotation curves of disk galaxies at large galactocentric radii
could be a peculiar effect of GR in rotating systems, instead of the
dynamical signature of the presence of DM halos. The suggestion is
somewhat puzzling, as observed galaxies are empirically in a
weak-field regime, with $v/c\approx 10^{-3}$, and so GR corrections
are expected to be very small: however, GR is a non-linear theory, and
so important effects cannot be excluded in principle.

In this paper, instead of attempting a GR modelization of the rotation
curve of some observed galaxy, we followed a different approach, and
we built the rotation curve for purely barionic disks, structurally
resembling real galaxies, and with total mass and scale lenght in the
observed range. The rotation curves are computed both in newtonian
gravity and in the (gravitomagnetic) weak field formulation of GR, and
compared.  After a detailed discussion of the mathematics, we
considered first the case of razor-thin disks made by purely circular
orbits. The orbital structure is highly idealized, but the surface
density of the disk is the standard exponential profile adopted to
model the stellar distribution in real disk galaxies; we also
considered another disk model, the Kuzmin-Toomre model, to gain
confidence with the results obtained for the exponential
disk. Consistently with the gravitomagnetic approximation adopted, we
impose ``self-consistency'' on the rotational velocity, i.e., we
assume that the rotational velocity of the test particle (star, or gas
cloud), and the rotational velocity of the stellar population
producing the gravitomagnetic field, are the same at each radius, and
we require that the rotational velocity is the solution of the
Lorentz-like equation of motion. The resulting problem is shown to be
a regular perturbation problem, with an intrinsinc order parameter
$\epsilon\approx 10^{-6}$, and the gravitomagnetic field is obtained
from two different (but equivalent) methods, i.e., by using elliptic
integrals and Bessel functions. It is found that the GR (normalized)
perturbative term is of the order of the unity, with no indications of
anomalously large emerging effects, and the resulting physical
modification of the newtonian rotation curve is therefore of the order
of $\epsilon\approx 10^{-6}$, well below the possibility to detection
in observed rotation curves, in agreement with the order-of-magnitude
estimate obtained by Toth (2021). Curiously, in the external regions
of the disk the gravitomagnetic GR perturbative term tend to {\it
  decrease} the rotational speed, as can be easily understood by
considering the gravitomagnetic field as given by the sum of the
fields produced by each current ring of the disk, and by the fact that
the disk current in {\it realistic} disk galaxies decreases
sufficiently fast at increasing radius, so that in the outer regions
of the disk the gravitomagnetic field is dominated by the inner
current distribution.  An important warning follows: if one assumes a
disk surface density distribution with a sufficiently slow radial
decline at increasing $R$, it is quite easy to show that the sum of
the fields of {\it external} rings can be very large (or even diverge)
leading to the wrong conclusion of a significant GR effect in real
galaxies: in theoretical/numerical studies of the present problem, the
use of {\it realistic} density profiles for the disks it is of
fundamental importance.

As doubts have been casted that disk thickness effects, or motions
more complicated than purely circular orbits for the stars producing
the gravitomagnetic fields, could be important, we also studied the
case of the gravitomagnetic rotation curve in genuinely tridimensional
stellar systmes. We first derived the gravitomagnetic Jeans equations
starting directly form the collisionless Boltzamnn equation, and the
associated Jeans theorem. In this case the equations contain naturally
a vertical ``pressure'' term due to the vertical velocity
dispersion. We showed how in axisymmetric systems the gravitomagnetic
field depends on the azimuthal streaming velocity field of the
system (even though the orbits of the single stars are in general {\it
  not} axisymmetric). In such a system, we can identity two different
rotational velocities: the circular velocity of a tracer in the
equatiorial plane, and the (circular) streaming velocity of the
stellar population producing the gravitomagnetic field. For both cases
we found that the GR perturbative term, after multiplication by
the expansion parameter $\epsilon\approx 10^{-6}$, is completely
negligible over the newtonian term, so that also for
collisionless axysimmetric systems the GR effects due to rotation
appear to be well below detectability, and unable to reproduce a flat
rotation curve at large radii any better than a newtonian model in
absence of DM.

In conclusion, the study conducted in the paper excludes the
possibility that gravitomagnetic GR effects can compensate by any
detectable amount the keplerian fall of the rotational velocity that
would characterize disk galaxies at large distances in absence of DM
halos, and produce the observed flat profiles: from the observational
point of view, in rotating disk galaxies DM is required by GR exactly
as in newtonian gravity. Of course, if a full GR simulation for the
same barionic disks used in this papers (with disk parameters
corresponding to the observed weak field regime of real galaxies)
convincingly proves the opposite, then among other things we should
conclude that gravitomagnetic weak field approximation of GR {\it
  cannot} be used to describe the weak field regime in rotating
galactic disks.

\acknowledgements I thank Giuseppe Bertin, Antonio Mancino, Bahram
Mashhoon, Jerry Ostriker, Silvia Pellegrini, Francesco Pegoraro, Renzo
Sancisi, Massimo Stiavelli, and an anonymous Referee, for important
comments. Alberto Parmeggiani (Department of Mathematics of Bologna
University), is especially thanked for discussions about some
mathematical aspect of this work.

\appendix

\section{Mathematical identities}

Here we list the most important mathematical identities used in this
work (see, e.g., BT08, C21, J98), and references therein).

The rotor of a vector function in cyclindrical coordinates is given by
\begin{equation}
\nabla\wedge\Av =\left({1\over R}{\partial\Az\over\partial\varphi} - {\partial\Aphi\over\partial z}\right)\eR +
                      \left({\partial\AR\over\partial z} -  {\partial\Az\over\partial R}\right)\ephi +
                       {1\over R}\left({\partial R\Aphi\over\partial R} -
                         {\partial\AR\over\partial\varphi}\right)\ez .
\label{eq:rotA}
\end{equation}

The complete elliptic integrals of first and second kind in Legendre
form are given respectively by
\begin{equation}
\begin{cases}  
\displaystyle{\Kc (k)=\int_0^{\pi/2}{d\varphi'\over\sqrt{1-k^2\sin^2\varphi'}} =
\int_0^1{d t\over\sqrt{(1-t^2)(1-k^2 t^2)}},}\cr\cr 
\displaystyle{\Ec (k)=\int_0^{\pi/2}\sqrt{1-k^2\sin^2\varphi'}d\varphi' 
=\int_0^1\sqrt{{1-k^2 t^2\over 1-t^2}}dt;}
\end{cases}
\label{eq:KEell}
\end{equation}
and from eq. (8.123) of GR07, 
\begin{equation}
\begin{cases}  
\displaystyle{{d\Kc (k)\over dk}={\Ec(k)\over k (1-k^2)}- {\Kc(k)\over k},}\cr\cr 
\displaystyle{{d\Ec (k)\over dk}={\Ec(k) -\Kc(k)\over k}.}
\end{cases}
\label{eq:difKEell}
\end{equation}
The two integrals over $\varphi'$ in Equations
(\ref{eq:Afield})-(\ref{eq:Bfield}) leading to Equations
(\ref{eq:Acylfull})-(\ref{eq:BfullK}), after reduction to the first
quadrant with the bisection formula $\cphip =2\cos^2(\varphi'/2) -1$,
become
\begin{equation}
\funZ = \int_0^{2\pi}{d\varphi'\over\sqrt{R^2+\xi^2 -2 R\xi\cphip +\Dz^2}} =
  {4 \Kc  (k)\over \sqrt{(R+\xi)^2+\Dz^2}},
\label{eq:int0}  
\end{equation}
\begin{equation}
\funU =\int_0^{2\pi}{\cphip\,d\varphi'\over\sqrt{R^2+\xi^2 -2 R\xi\cphip +\Dz^2}}
={4\over \sqrt{(R+\xi)^2+\Dz^2}}
\left[
  \left({2\over k^2}-1\right)\Kc(k) - {2\Ec(k)\over k^2}
\right], 
\label{eq:int1}
\end{equation}
where\footnote{Notice that in Mathematica, 
  $\Kc(k)={\rm EllipticK}[k^2]$, and   $\Ec(k)={\rm EllipticE}[k^2]$.}
\begin{equation}
  k^2 = {4 R\xi\over (R + \xi)^2+\Dz^2},\qquad\Dz=z-z'.
  \label{eq:kpara}
\end{equation}
For the considerations in Sections 2.1 and 2.2, we recall the
important identity
\begin{equation}
-{\partial\funZ\over\partial\xi}={1\over
  R}{\partial\, R\funU\over\partial R},
\label{eq:F0F1rot}
\end{equation}
that can be proved by differentiation of the explicit expressions in
Equations (\ref{eq:int0})-(\ref{eq:int1}), or (with some work) more
elegantly directly from their integral representation.

In numerical applications, when using Equations
(\ref{eq:Acylfull})-(\ref{eq:BfullK}), it is important to recall that
$\Ec(k)\sim 1$ for $k\to 1^{-}$, but
$\Kc(k) \sim - \ln\sqrt{1-k^2}\sim - \ln\sqrt{1-k}$, so that near the
ring $\xi = R$ in the $\Dz=0$ plane both $\funZ$ and $\funU$ diverge
as
\begin{equation}
\funZ\sim\funU\sim
  \begin{cases}
\displaystyle{-{4\ln\vert R-\xi\vert\over R+\xi},\qquad \Dz=0,\quad\xi\to R;}\cr\cr
 \displaystyle{-{2\ln\vert\Dz\vert\over R},\qquad\quad \xi=R,\quad\Dz\to 0};
\end{cases}
    \label{eq:int3}
\end{equation}
however the singularity is integrable and easily treated numerically.

From the theory of Green functions it is possible to prove that in
cylindrical coordinates (e.g., BT08, C21)
\begin{equation}
{1\over\Vert\xv -\yv\Vert}=\sum_{m=-\infty}^{\infty}{\rm e}^{i
  m(\varphi-\varphi')}\int_0^{\infty}{\rm e}^{-\lambda\vert\Dz\vert}\Jm(\lambda R)\Jm(\lambda\xi)\,d\lambda, 
\label{eq:greenB}  
\end{equation}
where $\Jm$ are the Bessel functions of first kind and integer order
$m$, $\xv=(R\cphi,R\sphi,z)$, $\yv=(\xi\cphip,\xi\sphip,z')$, and the
Hankel transform for a function $f(R)$ of the cylindrical radius
reads
\begin{equation}
\fHm (k)=\int_0^{\infty}R\,\Jm(\lambda R)\, f(R)\, dR,\qquad 
f(R)=\int_0^{\infty}\lambda\,\Jm(\lambda)\,\fHm (\lambda)\, d\lambda. 
\label{eq:hankm}  
\end{equation}

\section{Gravitomagnetic Jeans equations in cylindrical coordinates}

By using the standard approach of considering velocity moments of
Equation (\ref{eq:CBE}), and by working on cylindrical coordinates,
from the velocity moment of order 0 we obtain the continuity equation
\begin{equation}
  {\partial\rho\over\partial t}+{1\over R}{\partial
    R\rho\vR\over\partial R} + {1\over
    R}{\partial\rho\vphi\over\partial\varphi}
  +{\partial\rho\vz\over\partial z} = 0, 
\end{equation}
and from the three moments of order 1 (respectively over $\vpz$,
$\vpR$, and $\vpphi$) the three momentum equations
\begin{equation}
  \begin{cases}
    \displaystyle{
      {\partial\rho\vz\over\partial t}+
      {1\over R}{\partial R\rho\overline{\vpR\vpz}\over\partial R}+
     {1\over R}{\partial\rho\overline{\vpphi\vpz}\over\partial\varphi}+
     {\partial\rho\overline{\vpz^2}\over\partial z}
      =-\rho{\partial\phi\over\partial z} -\rho (\vR\Bphi -\vphi\BR),}\cr\cr 

     \displaystyle{
      {\partial\rho\vR\over\partial t}+
      {\partial\rho\overline{\vpR^2}\over\partial R}-
      {\rho (\overline{\vpphi^2}-\overline{\vpR^2})\over R}+
     {1\over R}{\partial\rho\overline{\vpR\vpphi}\over\partial\varphi}+
     {\partial\rho\overline{\vpR\vpz}\over\partial z}
      =-\rho{\partial\phi\over\partial R} -\rho (\vphi\Bz
      -\vz\Bphi),}\cr\cr

      \displaystyle{
      {\partial\rho\vphi\over\partial t}+
     {1\over R^2} {\partial R^2\rho\overline{\vpR\vphi}\over\partial R}+
     {1\over R}{\partial\rho\overline{\vpphi^2}\over\partial\varphi}+
     {\partial\rho\overline{\vpphi\vpz}\over\partial z}
      =-{\rho\over R}{\partial\phi\over\partial\varphi} -\rho (\vz\BR -\vR\Bz),}   
\end{cases}  
\label{eq:jeansCF}
\end{equation}
where the $\Bv$ field is determined by Equation (\ref{eq:BS}) computed
over the galaxy streaming mass density current given by the second of
Equation (\ref{eq:lorentzJ}). Finally, if the system is in steady
state, and supported by a phase-space DF $f(E,J_z - R\Aphi)$, from the
discussion in Section 4 it follows that Equation (B1) and the last of
Equations (B2) are identically verified, while the first two of
Equations (B2) reduce to Equation (\ref{eq:jeansEJ}).

\bigskip\bigskip
\twocolumngrid

\end{document}